\documentclass[aps,prd,preprint,superscriptaddress]{revtex4-1}
\usepackage{graphicx}
\usepackage{amsmath}
\usepackage{amssymb}
\usepackage{slashed}

\def\m0{m^{\!\!\!\!^o}}
\def\be{\begin{eqnarray}}
\def\ee{\end{eqnarray}}
\def\bc{\begin{center}}
\def\ec{\end{center}}

\def\om{\omega}

\def\Im{{\rm Im\,}}

\begin{document}

\title{Spectral representation for u- and t-channel exchange processes \\ in a partial-wave decomposition}


\author{M.F.M. Lutz}
\affiliation{GSI Helmholtzzentrum f\"ur Schwerionenforschung GmbH, \\Planckstra\ss e 1, 64291 Darmstadt, Germany}
\author{E.E. Kolomeitsev}
\affiliation{Matej Bel  University, Faculty of Natural Sciences, Tajovskeho 40, SK-97401 Banska Bystrica, Slovakia}
\author{C.L. Korpa}
\affiliation{Department of Theoretical Physics, University of P\'ecs,\\ Ifj\'us\'ag \'utja 6, 7624 P\'ecs, Hungary}
\date{\today}

\begin{abstract}
We study the analytic structure of partial-wave amplitudes derived from u- and t-channel exchange processes.
The latter plays a crucial role in dispersion-theory approaches to coupled-channel systems that model final state
interactions in QCD. A general spectral representation is established that is valid in the presence of anomalous thresholds,
decaying particles or overlapping left-hand and right-hand cut structures as it occurs frequently in hadron physics.
The results are exemplified at hand of ten specific processes.
\end{abstract}

\pacs{12.38.-t,12.38.Cy,12.39.Fe,12.38.Gc,14.20.-c ??}
\keywords{dispersion theory, anomalous threshold, partial-wave amplitudes }

\maketitle

\section{Introduction}

It is still an open challenge to derive final state interactions from QCD based on
effective field theory approaches at energies where the strong interaction forms resonances.
From the phenomenology of the last decades it is known that coupled-channel unitarity together
with the micro-causality condition play a decisive role in the enterprise to unravel the underlying
physics of this non-perturbative domain of QCD (see e.g. \cite{Zachariasen:1962zz,Dalitz:1967fp,Logan:1967zz,Kaiser:1995eg,GomezNicola:2001as,Lutz:2001yb,Lutz:2001mi,Kolomeitsev:2003kt,Lutz:2003fm,Gasparyan:2010xz,Danilkin:2010xd,Danilkin:2011fz,Gasparyan:2012km,Danilkin:2012ua}).

While final state interactions close to an elastic threshold can be treated quite reliably in perturbation theory
based on a suitable chiral Lagrangian this is not so for energies where the resonance spectrum
is observed. A convenient framework to study final state interactions is based on the concept of
a generalized potential. A partial-wave scattering amplitude  $T_{ab}(s)$ with a channel index $a$ and $b$
for the final and the initial state respectively is decomposed into contributions from left- and right-hand cuts where
all left-hand cut contributions reside in the generalized potential $U_{ab}(s)$. For an approximated generalized
potential the right-hand cuts are induced by means of the non-linear integral equation
\begin{eqnarray}
T_{ab}(s) = U_{ab}(s) +
\sum_{c,d}\int \frac{d w^2 }{\pi} \,\frac{s-\mu^2_M}{w^2-\mu^2_M}\,
\frac{T^\dagger_{ac}(w^2)\,\rho_{cd}(w^2)\,T_{db}(w^2)}{w^2-s-i\, \epsilon } \,,
\label{def-generalized-potential}
\end{eqnarray}
where $\rho_{cd}(w^2)$ is a channel dependent phase-space function.
By construction any solution of (\ref{def-generalized-potential}) does satisfy the coupled-channel
s-channel unitarity condition. While the general framework is known from the 60's of last
century \cite{Chew:1955zz,Mandelstam:1958xc,Chew:1961ev,Frye:1963zz,Ball:1969ri,Chen:1972rq,Eden:1966,Johnson:1979jy} only recently
this framework has been successfully integrated into an effective field theory approach based on the chiral
Lagrangian. The main additional and novel idea is to approximate the generalized potential systematically by means
of a conformal expansion that is reliable not only near threshold but also in the resonance region. The key observation
is that in (\ref{def-generalized-potential}) the generalized potential is needed only in the region where the partial-wave
amplitude has its right-hand cuts. In this region a conformal expansion is reliable and systematic
results can be expected. Since the expansion point for the conformal map can be dialed to lie within the convergence domain of
strict chiral perturbation theory the expansion coefficients may be computed from the chiral Lagrangian. First applications of
this novel approach can be found in \cite{Gasparyan:2010xz,Danilkin:2010xd,Danilkin:2011fz,Gasparyan:2011yw,Gasparyan:2012km,Danilkin:2012ua}.

The conformal expansion of the generalized potential requires the detailed knowledge of the spectral representation of
the generalized potential, the main target of the present work. The results of the following study are indispensable
for the analytic extrapolation of the generalized potential into the resonance region.  The analytic continuation of a function
requires a thorough understanding of its branch points and lines \cite{Kennedy-Spearman:1962}. The latter lead to its spectral representation.
While for reactions involving stable particles it is straight forward to unravel the spectral representation of the generalized
potential \cite{Petersen:1969pb,Kok:1973rb}, this is not so for reactions involving for instance the nonet of vector mesons with $J^P=1^-$
or the baryon decuplet states with $J^P = \frac{3}{2}^+$. The latter play a crucial role in the hadrogenesis conjecture that
expects the low-lying resonance spectrum of QCD-light with up, down and strange quarks only, to be generated by final state
interactions of the lowest SU(3) flavor multiplets with $J^P=0^-, 1^-$ and $J^P = \frac{1}{2}^+,  \frac{3}{2}^+$ \cite{Lutz:2001dr,Lutz:2001yb,Lutz:2001mi,
Kolomeitsev:2003kt,Lutz:2003fm,Lutz:2003jw,Kolomeitsev:2003ac,Lutz:2007sk,Hofmann:2005sw,Hofmann:2006qx,Terschlusen:2012xw}.
The coupled-channel interaction of such degrees of freedom leads to a plethora of subtle phenomena, which need to be treated carefully.
The left- and right-hand cuts may overlap and the generalized potential may be singular at threshold kinematics. The latter
leads to an anomalous threshold behavior of the partial-wave scattering amplitudes. This may occur at a threshold but also at a
pseudo-threshold. In this case the non-linear integral equation (\ref{def-generalized-potential}) has to be adapted properly.

The work is organized  as follows. In section II and III the framework for a dispersion-integral representation  
of partial-wave amplitudes is set up and general results are derived. 
Detailed illustrations are offered with specific t-channel and u-channel diagrams in section IV. 
We conclude with a  short summary in section V.

\newpage

\section{Partial-wave projection of invariant scattering amplitudes }

A general scattering amplitude $T(\bar k,k;w )$ will have a decomposition into a set of invariant amplitudes $F_n(s,t,u)$ and associated
tensors $L_n(\bar k,k;w )$ that carry possible Dirac and Lorentz structure of the scattering amplitude. The latter is required for reactions
of particles with non-vanishing spin. We write
\begin{eqnarray}
&&T(\bar k, \,k,\, w) = \sum_n \,F_n(s,t,u)\, L_n(\bar k,k;w )\,,
\qquad
\nonumber\\
&& s=(p+q)^2\,, \qquad t= (p-q)^2 \,,\qquad u = (p-\bar q)^2\,,
\end{eqnarray}
where we insist on invariant amplitudes, $F_n(s,t,u)$, that are free of kinematical constraints \cite{Trueman:1969tu,King:1970ud,Tindle:1975sd}. 
Owing to energy and momentum conservation
the scattering amplitude $T(\bar k,k;w )$ depends on three 4-vectors $\bar k_\mu,\,k_\mu$ and $w_\mu$ only with
\begin{eqnarray}
&& k={\textstyle{1\over 2}}\,(p-q)\,, \qquad \bar k
={\textstyle{1\over 2}}\,(\bar p-\bar q)  \,, \qquad w = p +q =\bar p+ \bar q\,,
\label{def-k}
\end{eqnarray}
where $p, q$ and $\bar p, \bar q$ are the 4-momenta of the in and outgoing particles respectively. 
A complete set of Dirac and Lorentz tensors $L_n(\bar k, k;w)$ depends on
the reaction considered. In the literature such a decomposition has been worked out explicitly for various reactions \cite{Bardeen:1969aw,King:1970ud,Cheung:1972tt,Stoica:2011cy,Lutz:2011xc,Heo:2014cja}.

The partial-wave scattering amplitudes are given by appropriate projection integrals
\begin{eqnarray}
T^{(JP)}(s) = \sum_n\int_{-1}^{+1} d x \,\lambda^{(JP)}_{n}(s\,,x)\,
F_n(s,t[s\,,x],u[s\,,x])\,,
\label{gen-rep-of-t}
\end{eqnarray}
where $\lambda^{(JP)}_n (s\,,x)$ are functions of kinematic origin. They are derived in the literature for any given angular momentum $J$
and parity $P$ (see e.g. \cite{Kibble:1960zz,Frazer:1960zz,Bardeen:1969aw,Cheung:1972tt,Stoica:2011cy,Lutz:2011xc,Heo:2014cja}). In (\ref{gen-rep-of-t}) 
we consider $F_n(s,t,u)$ as functions of $s$ and the cosine of the scattering angle $x= \cos \theta$. The main target of this work is the 
derivation of a spectral representation for such partial-wave amplitudes.

According to the hypothesis of Mandelstam \cite{Mandelstam:1958xc}, the amplitudes $F_n(s,t,u)$
satisfy dispersion integral representations characterized by a set of spectral
weight functions
\begin{eqnarray}
&& F_n(s,t,u) = \int_0^\infty \frac{d \bar s}{\pi} \,
\frac{\rho^{(n)}_{s}(\bar s)}{ s- \bar s}
+\int_0^\infty \frac{d \,\bar t}{\pi} \,
\frac{\,\rho^{(n)}_{t}(\bar t)}{t- \bar t}
+\int_0^\infty \frac{d \bar u}{\pi} \,
\frac{\rho^{(n)}_{u}(\bar u)}{u- \bar u}
\nonumber\\
&& \qquad  \qquad \;+\int_0^\infty \frac{d \bar s}{\pi} \int^\infty_0 \frac{d \,\bar t}{\pi} \,
\frac{\rho^{(n)}_{st}(\bar s, \bar t)}{( s-\bar s)\,(t- \bar t)}
+\int^\infty_0 \frac{d \,\bar t}{\pi} \int^\infty_0 \frac{d \bar u}{\pi} \,
\frac{\rho^{(n)}_{tu}(\bar t, \bar u)}{(t- \bar t)\,( u- \bar u)}
\nonumber\\
&& \qquad \qquad \;
+\int^\infty_0 \frac{d \bar s}{\pi} \int^\infty_0 \frac{d \bar u}{\pi} \,
\frac{\rho^{(n)}_{su}(\bar s, \bar u)}{(s-\bar s)\,( u- \bar u)} \,,
\label{rep-mandelstam}
\end{eqnarray}
as can be confirmed in perturbation theory. In effective field theory applications suitable subtractions may be required.
In this work we focus on the contributions defined by the t- and u-channel spectral weights $\rho^{(n)}_{t}(\bar t)$
and $\rho^{(n)}_{u}(\bar u)$. They give rise to so-called left-hand cuts in the partial-wave scattering amplitudes.
The s-channel contribution $\rho^{(n)}_{s}(\bar s)$ gives rise to s-channel unitarity cuts which are referred to as right-hand cuts.

\begin{figure*}[t]
\center{
\includegraphics[keepaspectratio,width=0.8\textwidth]{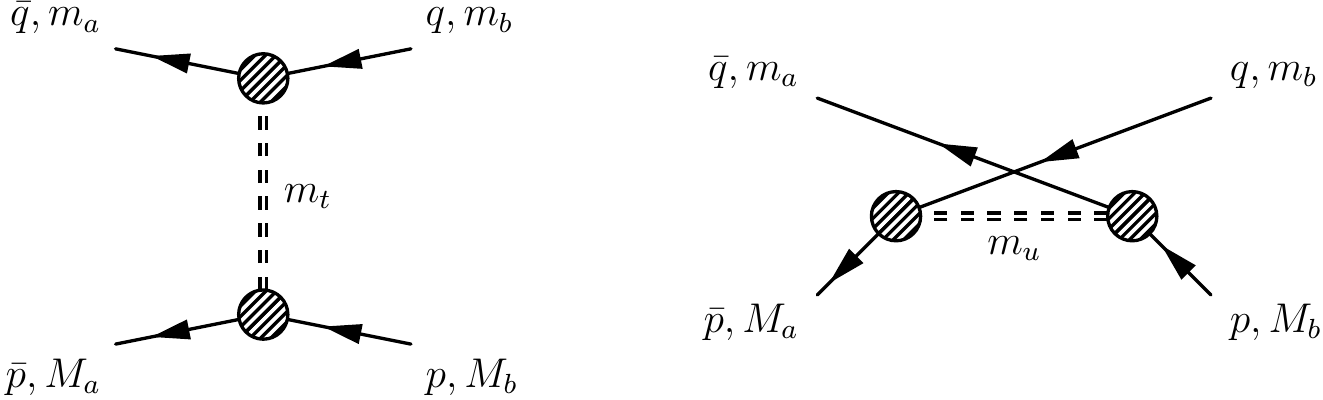} }
\caption{\label{fig:1} Generic t- and u-channel exchange processes.  }
\end{figure*}

In a first step we will establish a spectral representation for a generic t-channel and u-channel term as shown in Fig. \ref{fig:1}
\begin{eqnarray}
&& \int_{-1}^1 dx \,
\frac{\lambda_n(s,\,x)}{t[s\,,x]\,-m_t^2}
 = \sum_{i=\pm}\,
\int_{-\infty}^{\infty} \frac{d m^2}{\pi}\,\frac{\varrho^{(t)}_{n,i}(m^2,\,m_t^2)}{s-c^{(t\,)}_i(m^2)}\,
\left(\frac{d }{d m^2 }\,c^{(t\,)}_i(m^2) \right)\,,
\nonumber\\
&& \int_{-1}^1 dx \,
\frac{\lambda_n(s,\,x)}{u[s\,,x]-m_u^2}
 = \sum_{i=\pm}\,
\int_{-\infty}^{\infty} \frac{d m^2}{\pi}\,\frac{\varrho^{(u)}_{n,i}(m^2,\,m_u^2)}{s-c^{(u)}_i(m^2)}\,
\left(\frac{d }{d m^2 }\,c^{(u)}_i(m^2) \right)\,,
\label{disp-general-u-t-channel}
\end{eqnarray}
with the appropriate contour functions $c^{(t)}_\pm(m^2)$ and $c^{(u)}_\pm(m^2)$ that identify the location of the branch cuts and
some properly constructed spectral weights $\varrho^{(t)}_{n,\pm}(m^2,\,m_t^2)$ and $\varrho^{(u)}_{n,\pm}(m^2,\,m_u^2)$.
Given such a representations (\ref{disp-general-u-t-channel}) the general result for a partial-wave projected 
distributed t-channel or u-channel exchange as given in (\ref{rep-mandelstam}) in terms of $\rho^{(n)}_t(\bar t )$ and 
$\rho^{(n)}_u(\bar u )$ is readily obtained in terms of the folded spectral weights
\begin{eqnarray}
&& \varrho^{(t)}_{n,\pm}(m^2 ) = \int_0^\infty \frac{d \bar t}{\pi} \,\rho^{(n)}_t(\bar t ) \, \varrho^{(t)}_{n,\pm}(m^2,\,\bar t)\,, \qquad \quad
\varrho^{(u)}_{n,\pm}(m^2 ) =  \int_0^\infty \frac{d \bar u}{\pi} \, \rho^{(n)}_u(\bar u ) \, \varrho^{(u)}_{n,\pm}(m^2,\,\bar u) \,.
\label{def-effective-weight}
\end{eqnarray}
We note that a partial cancellation of the $+$ and $-$ contour contributions in (\ref{disp-general-u-t-channel}) may occur  
whenever the two contours run along identical regions on the real axis. 

While the derivation of the spectral weights $\varrho^{(t)}_{n,\pm}(m^2,\,t)$ and $\varrho^{(u)}_{n,\pm}(m^2,\,u)$
is quite cumbersome the identification of the contour functions $c^{(t)}_\pm(m^2)$ and $c^{(u)}_\pm(m^2)$
is straight forward. Owing to the Landau equations any possible branch point of a partial-wave amplitude
must be associated with an endpoint singularity of the projection integral (\ref{gen-rep-of-t}). Note that this is so only
if the invariant amplitudes $F_n(s,t,u)$ are free of kinematical constraints. In the presence of kinematical constraints
the functions $\lambda_n(s,\,x) $ may be singular at specific conditions which may lead to additional and unphysical branch points.
In our case the contour function may be introduced by the condition
\begin{eqnarray}
&& u[c_\pm^{(u)}(m^2)\,,\pm 1]=m^2 \,, \qquad \qquad t[c_\pm^{(t)}(m^2)\,,\pm 1]=m^2 \,.
\label{def-cis}
\end{eqnarray}

A few comments on the representation (\ref{disp-general-u-t-channel}) are in order. The integral on the left-hand
side of (\ref{disp-general-u-t-channel}) defines an analytic  function in $s$ with branch points at $s=0$
and $s = c^{(u)}_\pm(m_u)$.  Here we assume that the $x$-integration
contour  in (\ref{gen-rep-of-t}) is appropriately deformed into the complex plane to avoid the
situation $u[s,\,x] = m_u^2$ or $t[s,\,x] = m_t^2$ with $x \neq \pm 1$.  It is convenient, though not mandatory, to define
the branch cuts connected to the endpoint singularities of (\ref{disp-general-u-t-channel}), i.e. the points $c^{(t)}_\pm (m^2_t)$ or
$c^{(u)}_\pm (m^2_u)$, to lie on the lines defined by the functions $c^{(t)}_\pm(m^2)$ and $c^{(u)}_\pm(m^2)$. This procedure has the advantage
that t- and u-channel processes
with different exchange masses define branch cuts that are maximally overlapping. This is exploited in (\ref{def-effective-weight}).

The right-hand sides of (\ref{disp-general-u-t-channel}) may require a slight modification if the contour function
$c^{(t)}_\pm(m^2)$ or $c^{(u)}_\pm(m^2)$ hits a threshold point $s=(m_a\pm M_a)^2 $ or $(m_b \pm M_b)^2 $ at a critical value $m_{crit}$.
Such a need reflects the possible presence of an anomalous threshold \cite{Karplus:1958zz,Mandelstam:1960zz,Eden:1966}. In this case the 
contour has to be deformed close to $m_{crit}$. For instance, 
one may use a semicircle of radius $\epsilon$ centered around $m_{crit}$.

\newpage

\section{Spectral representation: general results }

The contour functions $c^{(t)}_{\pm}(m^2)$ and $c^{(u)}_{\pm}(m^2)$ depend on the masses of initial and final particles for which
we use the convenient notation
\begin{eqnarray}
q^2 = m_b^2 \,, \qquad \bar q^2 = m_a^2\,, \qquad p^2= M_b^2 \,,\qquad \bar p^2= M_a^2\,.
\end{eqnarray}
The root equations (\ref{def-cis}) for the contour functions can be solved analytically with the well known result
\begin{eqnarray}
&& c^{(u)}_{\pm,ab}(m^2)  = \frac{1}{2}\,\Big( M_a^2+m_a^2+M_b^2+m_b^2
-m^2\Big)
 +\frac{M_a^2-m_b^2}{\sqrt{2}\,m}\,
\frac{M_b^2-m_a^2}{\sqrt{2}\,m}
\nonumber\\
&&\pm\, \frac{m^2}{2}
\sqrt{ \Bigg(1-2\,\frac{M_a^2+m_b^2}{m^2}
+\frac{(M_a^2-m_b^2)^2}{m^4} \Bigg)\Bigg(
1-2\,\frac{M_b^2+m_a^2}{m^2}
+\frac{(M_b^2-m_a^2)^2}{m^4} \Bigg)}\,,
\nonumber\\
&&c^{(t)}_{\pm,ab}(m^2) =  \frac{1}{2}\,\Big(M_a^2+m_a^2+M_b^2+m_b^2-m^2 \Big)
- \frac{M_a^2-M_b^2}{\sqrt{2}\,m}\,
\frac{m_a^2-m_b^2}{\sqrt{2}\,m}
\nonumber\\
&&\pm \,\frac{m^2}{2} \sqrt{\Bigg( 1-2\,\frac{M_a^2+M_b^2}{m^2}
+\frac{(M_a^2-M_b^2)^2}{m^4} \Bigg) \Bigg(
1-2\,\frac{m_a^2+m_b^2}{m^2} +\frac{(m_a^2-m_b^2)^2}{m^4}\Bigg)}\,.
\label{lamtab}
\end{eqnarray}
The spectral weights  $\varrho^{(t)}_{n,\pm}(m^2,\,t)$ and $\varrho^{(u)}_{n,\pm}(m^2, \,u)$ introduced in (\ref{disp-general-u-t-channel})
factorize. This is a consequence of specific properties of the kinematical functions
$\lambda_n(s,\,x)$. They may have singularities at the thresholds $s=(m_a\pm M_a)^2 $ and
$(m_b \pm M_b)^2 $ only. However, the $x$-dependence in $\lambda_n(s,\,x)$  is such that the
integrals (\ref{disp-general-u-t-channel}) are finite at $s=(m_a\pm M_a)^2 $ and $(m_b \pm M_b)^2 $, at
least for sufficiently large $m_t$ and $m_u$. It holds
\begin{eqnarray}
&&\varrho^{(t)}_{n,\pm} (m^2,m_t^2) = \lambda_n(c^{(t)}_\pm(m^2),x^{(t)}_\pm (m^2)) \,\varrho^{(t)}_\pm(m^2,\,m_t^2),
\nonumber\\
&& x^{(t)}_\pm (m^2)= \frac{2\,\omega_{a}(s)\,\omega_{b}(s)-m_a^2-m_b^2+m_t^2
}{2\,p_{a}(s)\, p_{b}(s)} \,
\Bigg|_{s=c^{(t)}_\pm(m^2)}  \,,
\label{def-varrho-t}
\nonumber\\
&&\varrho^{(u)}_{n,\pm} (m^2,m_u^2) = \lambda_n(c^{(u)}_\pm(m^2),x^{(u)}_\pm (m^2)) \,\varrho^{(u)}_\pm(m^2,\,m_u^2),
\nonumber\\
&& x^{(u)}_\pm (m^2)= \frac{M_a^2+m_b^2-m_u^2-2\,E_{a}(s)\,\omega_{b}(s)
}{2\,p_{a}(s)\, p_{b}(s)} \,
\Bigg|_{s=c^{(u)}_\pm(m^2)}  \,,
\end{eqnarray}
with
\begin{eqnarray}
&& p_{i} (s) = \sqrt{\frac{(s-(m_i+M_i)^2)\,(s-(m_i-M_i)^2)}{4\,s}}\,,
\nonumber\\
&& \omega_{i}(s) = \frac{s-M_i^2+m_i^2}{2\,\sqrt{s}}\,,\qquad  \qquad
E_{i}(s) = \frac{s-m_i^2+M_i^2}{2\,\sqrt{s}}\,.
\label{def-relative-momenta}
\end{eqnarray}

The derivation of the master weight functions $\varrho^{(u)}_\pm(m^2,\,m_u^2)$ and $\varrho^{(t)}_\pm(m^2, \,m_t^2)$ is of utmost importance
for the present development  but quite tedious for the general case (see e.g. \cite{Stingl:1974im,Rinat:1971,Eden:1966}).
The authors did not find explicit expressions in the published literature for the general case. We present and
discuss first the simple case where the u-channel and t-channel exchange masses $m_u$ and $m_t$ are sufficiently large.
In this case the following results are readily established
\begin{eqnarray}
&& \varrho^{(t)}_\pm(m^2,\,m_t^2) =
\left\{
\begin{array}{l}
-\,\pi\,\frac{\Theta [m^2- m_t^2] }{2\,p_a(s)\,p_b(s)} \,\Big|_{s =c^{(t)}_\pm(m^2)} \qquad \quad {\rm for} \quad
{\rm Min}\{t^{(a)}_I,\,t^{(b)}_I\}  \leq 0 \\
\pm\,\pi\, \frac{\Theta [m^2- m_t^2] }{2\,p_a(s)\,p_b(s)} \,\Big|_{s =c^{(t)}_\pm(m^2)}\, \qquad \quad {\rm for}
\quad {\rm Min}\{t^{(a)}_I,\,t^{(b)}_I\}   > 0
\end{array}
\right. \,,
\nonumber\\
\nonumber\\
&& t^{(a)}_I= \frac{-M_a^2\,m_b^2 + m_a^2\,M_b^2}{m_a^2 - M_a^2}\,, \qquad
t^{(b)}_I= \frac{M_a^2\,m_b^2 - m_a^2\,M_b^2}{m_b^2 - M_b^2}\,.
\label{result-t-simple}
\end{eqnarray}
While the form of the spectral weight is quite simple and intuitive its associated phase factor
is complicated reflecting the choices of various Riemann sheets. We follow here a pragmatic approach. We will 
not give complicated arguments which Riemann sheets to choose, rather we present the final answer and assure 
that (\ref{disp-general-u-t-channel}) was
verified by numerical simulations for sufficiently large energies. It is worth pointing out that (\ref{result-t-simple})
holds for arbitrarily small exchange masses for the limiting  case $m_a=m_b$ and $M_a=M_b$ with $t_I^{(a)}=t_I^{(b)}=0$.

The particular values $t_I^{(a)}$ and $t_I^{(b)}$ are determined by the conditions
\begin{eqnarray}
&& \Im p^2_a\,c_\pm^{(t)}(m^2) = 0 \quad \& \quad \Im c_\pm^{(t)}(m^2) \neq 0 \qquad \to \qquad m^2 = t^{(a)}_I\,,
\nonumber\\
&& \Im p^2_b\,c_\pm^{(t)}(m^2) = 0 \quad \& \quad \Im c_\pm^{(t)}(m^2) \neq 0 \qquad \to \qquad m^2 = t^{(b)}_I\,,
\label{def-ti}
\end{eqnarray}
where we assure the independence of the solutions with respect to the contour indices $\pm$.

Before proceeding with a discussion of the more general case with an arbitrarily small
exchange mass $m_t$  we provide the analogous result for the u-channel term:
\begin{eqnarray}
&& \varrho^{(u)}_\pm(m^2,\,m_u^2) =
\left\{
\begin{array}{l}
-\,\pi\,\frac{\Theta [m^2- m_u^2] }{2\,p_a(s)\,p_b(s)} \,\Big|_{s =c^{(u)}_\pm(m^2)} \qquad {\rm for} \quad
{\rm Min} \{u^{(a)}_I,\,u^{(b)}_I\}  \leq 0 \\
\pm\, \pi\,\frac{\Theta [m^2- m_u^2] }{2\,p_a(s)\,p_b(s)} \,\Big|_{s =c^{(u)}_\pm(m^2)}\, \qquad {\rm for}
\quad {\rm Min} \{u^{(a)}_I,\,u^{(b)}_I\}  > 0
\end{array}
\right. \,,
\nonumber\\ \nonumber\\
&& u^{(a)}_I= \frac{-M_a^2\,M_b^2 + m_a^2\,m_b^2}{m_a^2 - M_a^2}\,, \qquad
u^{(b)}_I= \frac{-M_a^2\,M_b^2 + m_a^2\,m_b^2}{m_b^2 - M_b^2}\,.
\label{result-u-simple}
\end{eqnarray}
with
\begin{eqnarray}
&& \Im p^2_a\,c_\pm^{(u)}(m^2)= 0 \quad \& \quad \Im c_\pm^{(u)}(m^2) \neq 0 \qquad \to \qquad m^2 = u^{(a)}_I\,,
\nonumber\\
&& \Im p^2_b\,c_\pm^{(u)}(m^2) = 0 \quad \& \quad \Im c_\pm^{(u)}(m^2) \neq 0 \qquad \to \qquad m^2 = u^{(b)}_I\,.
\end{eqnarray}
In (\ref{result-u-simple}) we assume $m_u$ to be sufficiently large.  Note the formal similarity
of the expressions for the contour functions $c_\pm^{(u)}(m^2)$ and $c_\pm^{(t)}(m^2)$ as given in
(\ref{lamtab}): applying $m_b \leftrightarrow M_b$ transforms the expressions into each other.

We turn to the general case with arbitrary exchange masses. It suffices to provide explicit expressions for the
t-channel  case. Corresponding expressions valid for the u-channel term follow by the replacement $m_b \leftrightarrow M_b$.

In a first step we identify the points where a change of Riemann sheets and therewith a phase change  may be required.
All together there are 15  critical values for the squared exchange mass $m^2_t$. The expression (\ref{result-t-simple})
is valid for $m^2_t$ larger than the maximum of those 15 values.
Two points $t^{(a)}_I$ and $t_I^{(b)}$ we encountered already in  (\ref{result-t-simple}, \ref{def-ti}).
Additional four points are  determined by the conditions that the
squared contour functions pass through the threshold points $(m_a\pm M_a)^2 $ and
$ (m_b\pm M_b)^2$. It is intuitive that the latter are associated with a change of Riemann sheets and therefore will possibly
cause a phase change of the spectral weight at such points. We introduce
\begin{eqnarray}
&&  t^{(a)}_+= \frac{ m_a\,M_b^2+m_b^2\,M_a }{m_a + M_a} - m_a\,M_a\,,\qquad
t^{(b)}_+ =\frac{ m_b\,M_a^2+m_a^2\,M_b }{m_b + M_b} - m_b\,M_b \,,
\nonumber\\
&&c^{(t)}_\pm (m^2)  = (m_a+M_a)^2 \qquad  \to \qquad m^2 = t_+^{(a)}  \,,
\nonumber\\
&& c^{(t)}_\pm (m^2)  = (m_b+M_b)^2 \,\qquad  \to \qquad m^2 = t_+^{(b)}  \,,
\label{def-t-plus}
\end{eqnarray}
and
\begin{eqnarray}
&&  t^{(a)}_-= \frac{ m_a\,M_b^2-m_b^2\,M_a }{m_a - M_a} + m_a\,M_a\,,\qquad
t^{(b)}_- =\frac{ m_b\,M_a^2-m_a^2\,M_b }{m_b - M_b} + m_b\,M_b \,,
\nonumber\\
&&c^{(t)}_\pm (m^2) = (m_a-M_a)^2 \qquad  \to \qquad m^2 = t_-^{(a)}  \,,
\nonumber\\
&& c^{(t)}_\pm (m^2)  = (m_b-M_b)^2 \, \qquad  \to \qquad m^2 = t_-^{(b)}   \,,
\label{def-t-min}
\end{eqnarray}
where a solution exists either with respect to the subscript $\pm \to +$ or $\pm \to -$ depending on the specifics of
case. The $c^{(t)}_+(m^2)$  contour runs through two threshold points at most. The same holds for the
$c^{(t)}_-(m^2)$  contour: it may intersect the two threshold points that are avoided by $c^{(t)}_+(m^2)$.

Four further critical points are determined by the condition that the imaginary parts of the squared contour
functions approach zero: the argument of  the square root in (\ref{lamtab}) must vanish
\begin{eqnarray}
&& m^2_\pm = (m_a \pm m_b)^2 \,, \qquad \qquad \qquad M_\pm^2 = (M_a\pm M_b)^2 \,.
\nonumber\\
&& v_\pm^{+} ={\rm Max}\{m^{2}_\pm,\, M^{2}_\pm \} \,, \qquad \qquad \quad  \!\!
v_\pm^{-} ={\rm Min}\{m^{2}_\pm,\, M^{2}_\pm \}  \,.
\label{def-m-M-pm}
\end{eqnarray}
The critical values (\ref{def-m-M-pm}) determine whether the squared contour functions
lie on the real axis or invade the complex plane. The latter holds for
\begin{eqnarray}
\begin{array}{l}
v_+^{-}  < m^2 <  v_+^{+}   \\
v_-^{-}   < m^2 < v_-^{+}
\end{array}
\quad
\leftrightarrow  \quad \Im c_\pm^{(t)}(m^2) \neq 0  \qquad {\rm if} \qquad v_+^{-}  > v_-^{+}\,,
\label{property-v1}
\end{eqnarray}
and
\begin{eqnarray}
\begin{array}{l}
v_-^{+}   < m^2 < v_+^{+}  \\
v_-^{-}  \, < m^2 <  v_+^{-}
\end{array}
\quad
\leftrightarrow  \quad \Im c_\pm^{(t)}(m^2) \neq 0 \qquad {\rm if} \qquad v_+^{-}  < v_-^{+}\,.
\label{property-v2}
\end{eqnarray}
The points $m_+$ and $M_+$ have a direct physical interpretation: for $m_t > m_+ $ or $m_t> M_+$ the t-channel
exchange particle turns unstable.  Similarly the critical points $m_-$ and $M_-$ reflect the opening of decay channels
of initial or final particles.

In order to derive the generalization of (\ref{result-t-simple}) it is important to study the position of the
critical points $v^{+}_\pm$ and $v^-_\pm$ in relation to the points $t^{(a)}_I$ and $t_I^{(b)}$ introduced already in
(\ref{result-t-simple}).  We derive the inequalities
\begin{eqnarray}
&&  v_-^- < t_I^- < v_+^-  \quad \& \quad t^+_I > v_-^+ \qquad \quad {\rm for} \qquad t_I^- >0 \,,
\nonumber\\
&& t^+_I < v_+^-  \qquad   \qquad \qquad \qquad \qquad \quad \;\, \, {\rm for} \qquad t_I^- <0  \,,
\nonumber\\
&& t_I^{+} \,={\rm Max}\{t^{(a)}_I, \,t^{(b)}_I \} \,, \qquad \qquad \!\!\!
t_I^{-}\, ={\rm Min}\{t^{(a)}_I,\, t^{(b)}_I \}  \,,
\label{ineq}
\end{eqnarray}
which follow with ease from the two identities
\begin{eqnarray}
&& t_I^{(a)} = M_b^2+ \frac{t_I^{(a)}}{t_I^{(b)}}\,M_a^2 = m_b^2+ \frac{t_I^{(a)}}{t_I^{(b)}}\,m_a^2 \,,
\nonumber\\
&& t_I^{(b)} = M_a^2+ \frac{t_I^{(b)}}{t_I^{(a)}}\,M_b^2 = m_a^2+ \frac{t_I^{(b)}}{t_I^{(a)}}\,m_b^2 \,.
\end{eqnarray}
It is useful to work out also the relative positions of the remaining critical points.
After tedious considerations we find the relations
\begin{eqnarray}
&& t_+^-\leq v_-^- \leq {\rm Min}\{v_-^+,  v_+^- \}\leq {\rm Min}\{ t_+^+, t_-^-\}
\nonumber\\
&& \qquad \quad \;\;\,  \leq {\rm Max}\{t_+^+, t_-^- \}\leq {\rm Max}\{v_-^+,  v_+^- \}
\nonumber\\
&& \qquad \quad \;\;\, \leq v_+^+ \leq t_-^+  \qquad \quad  {\rm for} \quad t_-^- > {\rm Min}\{v_-^+,  v_+^- \} \,,
\nonumber\\ \label{result-relative-positions} \\
&&  t_+^-\leq v_-^- \leq {\rm Min}\{v_-^+,  v_+^- \}\leq {\rm Min}\{ t_+^+, t_-^+\}
\nonumber\\
&& \qquad \quad \;\;\,  \leq {\rm Max}\{t_+^+, t_-^+ \}\leq {\rm Max}\{v_-^+,  v_+^- \}
\nonumber\\
&& \qquad \quad \;\;\,\leq v_+^+  \qquad \qquad \quad \;\,   {\rm for} \quad t_-^- \leq {\rm Min}\{v_-^+,  v_+^- \} \,,
\nonumber
\end{eqnarray}
where we introduced the notation
\begin{eqnarray}
&& t_\pm^{+} \,={\rm Max}\{t^{(a)}_\pm,\, t^{(b)}_\pm\} \,, \qquad \qquad
t_\pm^{-}\, ={\rm Min}\{t^{(a)}_\pm,\, t^{(b)}_\pm \}  \,.
\label{def-convention}
\end{eqnarray}

We are now prepared to display the master spectral weight, where we assume
$m_a\neq m_b$ or $M_a\neq M_b$ in the following. Recall that for the diagonal limit with
$m_a=m_b$ and $M_a=M_b$ the spectral weight is given by (\ref{result-t-simple}).
We discriminate four different cases
\allowdisplaybreaks[1]
\begin{eqnarray}
&& \varrho^{(t)}_{\pm}(m^2,\,m_t^2) =
\left\{
\begin{array}{l}
\varrho^{(t)}_{\pm,1}(m^2,\,m_t^2) \qquad {\rm for } \qquad    \; t^{-}_I  \leq  0    \quad  \& \quad
v_-^+  < v_+^- \\
\varrho^{(t)}_{\pm,2}(m^2,\,m_t^2)\qquad {\rm for } \qquad     \; t^{-}_I  > 0   \quad  \& \quad
v_-^+  < v_+^- \\
\varrho^{(t)}_{\pm,3}(m^2,\,m_t^2)\qquad {\rm for } \qquad      \; t^{-}_I  \leq  0     \quad  \& \quad
v_-^+  \geq  v_+^- \\
\varrho^{(t)}_{\pm,4}(m^2,\,m_t^2)\qquad {\rm for } \qquad     \; t^{-}_I   > 0    \quad  \& \quad
v_-^+  \geq  v_+^-
\end{array} \right.
\,,
\nonumber\\ \nonumber\\
&& \varrho^{(t)}_{\pm,i}(m^2,\,m_t^2) = \varrho^{(t)}_{\pm,i}(m)\,\Theta[m^2-m_t^2] \,,
\label{def-four-cases}
\end{eqnarray}
with
\begin{eqnarray}
&& \varrho^{(t)}_{\pm,1}(m^2) =\frac{\pi }{2\,p_a(s)\,p_b(s)} \,\Bigg|_{s =c^{(t)}_\pm(m^2)}\,
\Big\{ -  1
\nonumber\\
&& \qquad   + \,2\,\Theta [v^+_+-m^2]- 2\,\Theta [v^-_+-m^2]
 \nonumber\\
&& \qquad   -\,2\, \Theta [v^+_--m^2]\,\Theta[v^+_- -t^{+}_I]+2\, \Theta [v^-_--m^2]\,\Theta[v^-_- -t^{+}_I]
 \nonumber\\
&& \qquad +\,2\, \Theta [t^{+}_I-m^2]\,\Theta[t^{+}_I-v^-_- ]\,\Theta[v^+_- -t^{+}_I]
\nonumber\\
&& \qquad + \,2\,\Theta [t^{+}_\pm -m^2]\,-\, 2\,\Theta [t^{-}_\mp -m^2]
\Big\}\,,
\nonumber\\
&& \varrho^{(t)}_{\pm,2}(m^2) =\frac{\pi }{2\,p_a(s)\,p_b(s)} \,\Bigg|_{s =c^{(t)}_\pm(m^2)}\,
\Big\{ \pm  1
\nonumber\\
&& \qquad   + \,2\,\Theta [v^+_+-m^2]\,\Theta[t^{+}_I-v^+_+]
- 2\,\Theta [v^-_+-m^2] \,\Theta[t^{+}_I-v^-_+]
\nonumber\\
&& \qquad + \,2\,\Theta [t^{+}_I-m^2]\,\Theta[t^{+}_I-v^-_+]\,\Theta[v^+_+-t^{+}_I]
\nonumber\\
&& \qquad   - \,2\,\Theta [v^+_--m^2]\,\Theta[v^+_--t^{-}_I]
+ 2\,\Theta [v^-_--m^2] \,\Theta[v^-_--t^{-}_I]
\nonumber\\
&& \qquad + \,2\,\Theta [t^{-}_I-m^2]\,\Theta[t^{-}_I-v^-_-]\,\Theta[v^+_--t^{-}_I]
\nonumber\\
&& \qquad
+\, 2\,\Theta [t^{\pm}_\pm-m^2]- 2\,\Theta [t^{\pm}_\mp-m^2]
\Big\}\,,
\nonumber\\
&& \varrho^{(t)}_{\pm,3}(m^2) =\frac{\pi }{2\,p_a(s)\,p_b(s)} \,\Bigg|_{s =c^{(t)}_\pm(m^2)}\,
\Big\{ -  1
\nonumber\\
&& \qquad   + \,2\,\Theta [v^+_+-m^2] +\, 2\,\Theta [v^-_--m^2]\,\Theta[v^-_--t^{+}_I]
 \nonumber\\
&& \qquad + \,2\,\Theta [t^{+}_I-m^2]\,\Theta[t^{+}_I-v^-_-]
 -2\,\Theta [t^{+}_\mp -m^2]- 2\,\Theta [t^{-}_\mp -m^2]
\Big\}\,,
\nonumber\\
&& \varrho^{(t)}_{\pm,4}(m^2) =\frac{\pi}{2\,p_a(s)\,p_b(s)} \,\Bigg|_{s =c^{(t)}_\pm(m^2)}\,
\Big\{ \pm  1
\nonumber\\
&& \qquad   + \,2\,\Theta [v^+_+-m^2]\,\Theta [t_I^+- v^+_+]
+ 2\,\Theta [t^+_I-m^2]\,\Theta [v^+_+-t_I^+]
\nonumber\\
&& \qquad   + \,2\,\Theta [t^-_I-m^2]\,\Theta [t_I^--v^-_-]
 - 2\,\Theta [t^{+}_\mp -m^2] - 2\,\Theta [t^{-}_\mp-m^2]
\Big\}\,,
\label{central-result}
\end{eqnarray}
where we apply the convenient notations (\ref{def-m-M-pm}, \ref{ineq}, \ref{def-convention}).
The result (\ref{central-result}) deserves some discussion.  The first term in each of the
four expressions in (\ref{central-result}) describes the opening of a normal left-hand cut
at $m^2> m^2_t$.  While the conditions $m^2> v_-^\pm$
characterize the opening of decay channels of the exchanged particle the conditions
$m^2 < v_+^\pm$ signal an unstable initial or final state. Anomalous thresholds open at
$m^2 < t_\pm^+$ or $m^2 < t_\pm^-$. With (\ref{central-result}) we also specify implicitly which
contour runs through which threshold points. This follows since
each threshold point is associated with a sign change of the spectral functions as  detailed in
(\ref{central-result}) at $m^2 = t_\pm^+$ or $m^2=t_\pm^-$. For a given plus or minus contour and a selected
case $i=1,...,4$ in (\ref{def-four-cases}) two critical values out of the four $t_\pm^+$  and $t_\pm^-$ points are
selected unambiguously.

The merit of (\ref{central-result}) lies in its generality. It is a convenient starting point for coupled-channel studies
with many channels involved, where a case-by-case study is prohibitive. In certain cases the result (\ref{central-result}) may be
further simplified by the observation that there may be partial cancellations of the plus and minus contour contributions in
regions where they are moving on the real axis. With (\ref{central-result}) it is straight forward to implement such cancellations
in a computer code.

We alert the reader that the result (\ref{central-result}) requires an analytic continuation for the case that
a channel with non-zero angular momentum $L\neq 0$ is considered. This implies that the functions
$\lambda_n(s,\,x)$ in (\ref{disp-general-u-t-channel}) are singular at thresholds and consequently the
contour function $c^{(t)}_\pm(m^2)$ needs to be deformed into the complex $m^2$-plane close to the critical values 
$m^2_{crit}= t_\pm^+$ and $t_\pm^-$. Using semicircles centered around the critical points
this is readily achieved. The spectral weight is continued onto the deformed contour by the condition that it
is continuous along the semicircles. This leads to an unambiguous definition of the $\Theta$-functions in (\ref{central-result})
along the deformed contour: $\Theta$ functions in the phase parameters of the semi circles arise. Since the analytic expression
for the critical phases are quite complicated and implicit they may be determined numerically.

The spectral weights in (\ref{central-result}) are constructed such that
the representation (\ref{disp-general-u-t-channel}) holds for sufficiently large $s$. It does not necessarily hold
for arbitrarily small energies. For the specific case with $ m_t^2 < t_+^-$ the contour
function cuts through the larger threshold point
\begin{eqnarray}
c^{(t)}_{-,ab}(m^2) = {\rm Max }\{ (m_a+M_a)^2,\, (m_b+M_b)^2\}   \qquad {\rm with}\qquad  m^2 =t_+^- \,,
\label{}
\end{eqnarray}
and (\ref{disp-general-u-t-channel}) is not realized for energies slightly above that larger threshold. An analytic continuation
of the r.h.s. of (\ref{disp-general-u-t-channel}) is possible to affirm the realization of (\ref{disp-general-u-t-channel}) at the
larger threshold and above. We specify the analytic continuation by additional terms $\Delta \varrho^{(t-)}_{\pm,i}(m^2\,m_t^2)$ in
(\ref{central-result}). Replacing the spectral weight in (\ref{def-varrho-t}, \ref{def-four-cases}) by
\begin{eqnarray}
 \varrho^{(t)}_{\pm,i}(m^2,\,m_t^2)=\varrho^{(t)}_{\pm,i}(m^2)\,\Theta[m^2-m_t^2] + \Delta \varrho^{(t-)}_{\pm,i}(m^2,\,m_t^2)\,,
\label{def-analytic-continuation-generic}
\end{eqnarray}
will insure the validity of (\ref{disp-general-u-t-channel}) for energies exceeding the larger threshold point.

The construction of
$\Delta \varrho^{(t-)}_{\pm,i}(m^2,\,m_t^2)$ requires a  further set of specific contour points as conveniently
introduced by the condition
\begin{eqnarray}
&&c_\pm^{(t)} ( (m_a \pm m_b)^2) =c_\pm^{(t)} (\bar m^2_\pm ) \,, \qquad
c_\pm^{(t)} ((M_a \pm M_b)^2) =c_\pm^{(t)} (\bar M^2_\pm ) \,,
\label{def-barm-barM}
\end{eqnarray}
where we are interested in the solutions with $\bar m_\pm^2 \neq (m_a \pm m_b)^2 $ and $\bar M_\pm^2 \neq (M_a \pm M_b)^2 $.
The latter determine exchange masses where the contour function returns to itself.
We derive the explicit formulae
\allowdisplaybreaks[1]
\begin{eqnarray}
&& \bar m_\pm^2 = m_a^2 \mp \frac{m_b}{m_a}\,M_a^2+m_b\,\Big(m_b \mp m_a \Big)\,\frac{M_a^2-M_b^2}{m_a^2-m_b^2}
\nonumber\\
&& \qquad - \,\frac{m_b}{m_a}\,\frac{(m_b \pm m_a)\,(m_a^2-M_a^2)^2}{m_b\,(m_a\,m_b \pm (m_a^2 - M_a^2))-m_a\,M_b^2} \,,
\nonumber\\
&& \bar M_\pm^2 = M_a^2 \mp \frac{M_b}{M_a}\,m_a^2+M_b\,\Big(M_b \mp M_a \Big)\,\frac{m_a^2-m_b^2}{M_a^2-M_b^2}
\nonumber\\
&& \qquad - \,\frac{M_b}{M_a}\,\frac{(M_b \pm M_a)\,(M_a^2-m_a^2)^2}{M_b\,(M_a\,M_b \pm (M_a^2 - m_a^2))-M_a\,m_b^2} \,,
\nonumber\\ \nonumber\\
&& \{\bar v_\pm^{-}, \bar v_\pm^+\} =  \left\{
\begin{array}{ll}
\{\bar m_\pm ^2,\, \bar M_\pm^2\} \qquad & {\rm if} \quad (m_a\pm m_b)^2 < (M_a\pm M_b)^2 \\
\{\bar M_\pm ^2,\,\bar m_\pm^2\} \qquad & {\rm if} \quad (m_a\pm m_b)^2 > (M_a\pm M_b)^2
\end{array} \right. \,,
\label{def-barv}
\end{eqnarray}
and introduce further notations $\bar v_\pm^-$ and $\bar v_\pm^+$. While the points $v_\pm^+$ and $v_\pm^-$ characterize the
exchange masses where the contour leaves the real axis and invades the complex plane, the associated points $\bar v_\pm^-$ and $\bar v_\pm^+$
determine where the contour returns to those exit points possibly.
It is important to know the relative locations of the points $\bar v_\pm^+$ and $\bar v_\pm^-$ with respect to the
critical points introduced before.

We first focus on the relevant case with $t_+^- >0$ for which we derive the following
relations
\begin{eqnarray}
&& t_+^- >0 \qquad \to \qquad  \bar v_-^- \leq t_+^- \leq v_-^- \qquad \qquad \quad \;\,\; {\rm always} \,,
\label{closed-contour-conditions}\\ \nonumber\\
&&  t^-_-\leq 0\;\;\;\;\& \;\;
v_-^+ \leq t_+^+ \leq \bar v_-^+ \leq {\rm Min} \{v_+^-,\,t^+_- \} \qquad {\rm if} \quad t_I^- \leq  0   \;\; \& \; \;v_-^+ <  v_+^- \,,
\nonumber\\
&&  v_-^+ \leq t_+^+ \leq \bar v_-^+ \leq {\rm Min} \{v_+^-,\,t^-_- \} \leq t^+_- \qquad \qquad \quad  {\rm if} \quad t_I^- >  0 \;\; \& \;\; v_-^+ <  v_+^- \,,
\nonumber\\
&&   t^-_-\leq 0\;\;\;\;\& \;\;
v_+^- \leq t_-^+ \leq \bar v_+^- \leq {\rm Min} \{v_-^+,\,t_+^+\} \qquad {\rm if} \quad t_I^- \leq  0 \;\; \& \; \;v_-^+ \geq v_+^-  \,,
\nonumber\\
&&
v_+^- \leq t_-^- \leq \bar v_+^- \leq {\rm Min} \{v_-^+ ,\,t_+^+\}\leq t^+_- \qquad \qquad \quad {\rm if} \quad t_I^- > 0   \;\;\& \;\; v_-^+ \geq v_+^-\,.
\nonumber
\end{eqnarray}
The analytic continuation of the r.h.s. of (\ref{disp-general-u-t-channel}) is introduced upon the identification of an appropriate closed
contour, inside which the spectral weight is analytic. For an energy outside that closed domain the dispersion integral of
(\ref{disp-general-u-t-channel}), considered with respect to that closed contour, is unchanged.
For $s$ inside the closed domain it is altered necessarily. The closed contour
needed for the desired analytic continuation is readily identified with
\begin{eqnarray}
&& c^{(t)}_{-,ab}(m^2) \qquad {\rm with} \quad   \bar v_-^- < m^2 < \left\{
\begin{array}{ll}
v_-^+  \qquad & {\rm if} \qquad v_-^+ <v_+^- \\
v_+^- \qquad & {\rm if} \qquad v_+^-< v_-^+
\end{array} \right.\,,
\nonumber\\
&& c^{(t)}_{+,ab}(m^2) \qquad {\rm with} \quad   v_-^- < m^2 <
\left\{
\begin{array}{ll}
\bar v_-^+  \qquad & {\rm if} \qquad v_-^+ <v_+^- \\
\bar v_+^- \qquad &{\rm if} \qquad v_+^-< v_-^+
\end{array} \right.\,,
\label{def-closed-contour}
\end{eqnarray}
where a closed domain arises upon the union of the plus and minus contour lines introduced in (\ref{def-closed-contour}).
The spectral weights $\Delta \varrho^{(t-)}_{\pm,i}(m^2,\,m_t^2)$  in (\ref{def-analytic-continuation-generic}) follow from the 
condition that  the expressions 
(\ref{def-analytic-continuation-generic}) vanish for exchange masses $m > m_t$ on the closed contour as introduced in 
(\ref{def-closed-contour}). If the contour cuts through the largest threshold point and the spectral weight would be non-vanishing 
in this region a singular threshold behavior would arise necessarily from the r.h.s. of (\ref{disp-general-u-t-channel}). 
This would contradict the l.h.s. of (\ref{disp-general-u-t-channel}), which implies a regular behavior at the largest threshold point always. 
The situation is reconciled by the analytic continuation we are after. We find the result
\begin{eqnarray}
&& \Delta \varrho^{(t-)}_{\pm,1}(m^2,\,m_t^2) =
\pi\,\frac{\Theta [t_+^--m_t^2]  }{2\,p_a(s)\,p_b(s)} \,\Bigg|_{s =c^{(t)}_\pm(m^2)}
\nonumber\\
&& \qquad \,\times \Bigg\{ \left(2\,\Theta[m^2-t_I^+]\,\Theta[v^+_--t_I^+]  -1\right) \Theta[m^2-v_-^-]\,\Theta[v_-^+-m^2]
\nonumber\\
&& \qquad \;\;\,\,+ \left(1-2\,\Theta[\pm\, t^\pm_+ \,\mp\,m^2] \right)
\Theta[\pm \,\bar v_-^\pm \mp m^2]\,\Theta[\pm \,m^2 \mp v_-^\pm]\Bigg\}\,,
\nonumber\\
&& \Delta \varrho^{(t-)}_{\pm,2}(m^2,\,m_t^2) =
 \pi\,\frac{\Theta [t_+^--m_t^2]  }{2\,p_a(s)\,p_b(s)} \,\Bigg|_{s =c^{(t)}_\pm(m^2)}
\nonumber\\
&& \qquad \,\times \Bigg\{ \left(2\,\Theta[m^2-t_I^-]\,\Theta[v^+_--t_I^-]-1  \right)
 \Theta[m^2-v_-^-]\,\Theta[v_-^+-m^2]
\nonumber\\
&& \qquad \;\;\,\,+ \left(1-2\,\Theta[\pm\, t^\pm_+ \,\mp\,m^2] \right)
 \Theta[\pm \,\bar v_-^\pm \mp m^2]\,\Theta[\pm \,m^2 \mp v_-^\pm]
\Bigg\}\,,
\nonumber\\
&& \Delta \varrho^{(t-)}_{\pm,3}(m^2,\,m_t^2)= \pi\,\frac{ \Theta [t_+^--m_t^2]}{2\,p_a(s)\,p_b(s)} \,
\Bigg|_{s =c^{(t)}_\pm(m^2)}
\nonumber\\
&& \qquad \,\times \Bigg\{
 \left(1-2\,\Theta[t_I^+-m^2] \right)
  \Theta[m^2-v_-^-]\,\Theta[v_+^--m^2]
\nonumber\\
&& \qquad \;\;\,\,+\left(1-2\,\Theta [m^2-t^\pm_\mp] \right)  \Theta[\pm \,\bar v_\pm^- \mp m^2]\,\Theta[\pm \,m^2 \mp v_\pm^-]
\Bigg\}\,,
\nonumber\\
&& \Delta \varrho^{(t-)}_{\pm,4}(m^2,\,m_t^2) = \pi\,\frac{\Theta [t_+^--m_t^2]}{2\,p_a(s)\,p_b(s)} \,
\Bigg|_{s =c^{(t)}_\pm(m^2)}
\nonumber\\
&& \qquad \,\times \Bigg\{ \left( 1-2\,\Theta[t_I^- -m^2] \right)
\Theta[m^2-v_-^-]\,\Theta[v_+^--m^2]
\nonumber\\
&& \qquad \;\;\,\,+\left(1-2\,\Theta [m^2-t^-_\mp]  \right)  \Theta[\pm \,\bar v_\pm^- \mp m^2]\,\Theta[\pm \,m^2 \mp v_\pm^-]
\Bigg\}\,.
\label{analytic-continuation-result-a}
\end{eqnarray}
Due to the particular construction of $\Delta \varrho^{(t-)}_{\pm,i}(m^2,\,m_t^2)$ it is possible to write the total spectral
weights  in (\ref{def-analytic-continuation-generic}) directly in terms of the functions
$\varrho^{(t)}_{\pm,i}(m)$ introduced in (\ref{central-result}).
All together we affirm that using either (\ref{analytic-continuation-result-a}) in (\ref{def-analytic-continuation-generic}) or
\begin{eqnarray}
&&\varrho^{(t)}_{+,i}(m^2,\,m_t^2) = \Big\{- \Theta [t_+^--m_t^2] \, \Theta[m^2-v_-^-]\,\Theta[ \bar v^+_ -  -m^2]
\nonumber\\
&& \qquad \qquad \qquad \quad \;+\,\Theta[m^2-m_t^2]  \Big\} \,
\varrho^{(t)}_{+,i}(m) \qquad \qquad {\rm for} \;\; i=1,2\,,
\nonumber\\
&&\varrho^{(t)}_{+,i}(m^2,\,m_t^2) = \Big\{- \Theta [t_+^--m_t^2] \, \Theta[m^2-v_-^-]\,\Theta[ \bar v^-_+  -m^2]
\nonumber\\
&& \qquad \qquad \qquad \quad \;+\,\Theta[m^2-m_t^2]  \Big\} \,
\varrho^{(t)}_{+,i}(m)  \qquad \qquad {\rm for} \;\; i=3,4\,,
\nonumber\\
&& \varrho^{(t)}_{-,i}(m^2,\,m_t^2) = \Big\{-\Theta [t_+^--m_t^2] \, \Theta[m^2-\bar v_-^-]\,\Theta[\,{\rm Min}\{v^+_-,\,v^-_+ \} -m^2]
\nonumber\\
&&  \qquad \qquad \qquad \quad \; + \,\Theta[m^2-m_t^2] \Big\} \,
\varrho^{(t)}_{-,i}(m) \qquad \qquad {\rm for} \;\; i=1,2,3,4\,,
\label{res-together}
\end{eqnarray}
the validity of (\ref{disp-general-u-t-channel}) for energies exceeding the larger threshold point is ensured.

There is a further complication to be addressed.
The representation (\ref{disp-general-u-t-channel}) is not necessarily valid for
energies in between the two normal thresholds
\begin{eqnarray}
{\rm Min}\{m_a+M_a,\,m_b+M_b\}< \sqrt{s} < {\rm Max}\{m_a+M_a,\,m_b+M_b\} \,.
\end{eqnarray}
Two cases need to be discriminated. Either both pseudo-threshold values, $|m_a-M_a|$  and
$|m_b-M_b|$ are smaller than the two normal thresholds $m_a+M_a$ and $m_b+M_b $ or this is not true. In both cases
an analytic continuation of the l.h.s. of (\ref{disp-general-u-t-channel}), may be required. For the case of an inverted threshold order
with e.g.
\begin{eqnarray}
|m_b-M_b| \leq m_b+M_b \leq |m_a-M_a| \leq  m_a+M_a \,,
\label{def-case}
\end{eqnarray}
also the r.h.s. of (\ref{disp-general-u-t-channel}) needs an analytic
continuation for energies below the larger pseudo-threshold energy. Such an inversion occurs always for $i= 3 $ or
$i=4$  with $v^+_- \geq v^-_+$ but is impossible for $i=1$ or $i=2$ in (\ref{def-four-cases}).

We first construct the analytic continuation of the l.h.s. (\ref{disp-general-u-t-channel}), which is necessary
provided that the following condition is realized
\begin{eqnarray}
&& {\rm Max}\{ (m_a-M_a)^2, \,(m_b+M_b)^2 \} < s_+ (m_t^2) < (m_a+M_a)^2  \qquad
\nonumber\\
\!\!\!{\rm or} \;\; && {\rm Max}\{ (m_a-M_a)^2, \,(m_b+M_b)^2 \} < s_-(m_t^2) < (m_a+M_a)^2 \,,
\nonumber\\ \nonumber\\
&& s_\pm (m^2_t) = \frac{m_a^2+M_a^2+m_b^2+M_b^2}{2} - m_t^2
\nonumber\\
&& \quad \pm  \sqrt{\left(\frac{m_a^2+M_a^2+m_b^2+M_b^2}{2}-m_t^2\right)^2-(M_a^2-m_a^2)\,(M_b^2-m_b^2)} \,,
\label{def-spm}
\end{eqnarray}
where we assumed $m_a+M_a \geq  m_b+M_b$ without loss of generality. The particular functions $s_\pm (m^2_t)$ introduced in
(\ref{def-spm}) pass through
the thresholds at  the  critical points $m_t^2=t_\pm^{(a)}$ and $m_t^2=t_\pm^{(b)}$  introduced
in (\ref{def-t-plus}, \ref{def-t-min}) when studying the contour properties.
We derive
\begin{eqnarray}
&& {\rm for } \qquad t_\pm^{(a)} \geq \frac{m_b^2+M_b^2}{2}-\frac{m_a^2 + M_a^2}{2} \mp 2\,m_a\,M_a  \,,
\nonumber\\
&&  s_- (t_\pm^{(a)}) = (m_b^2 - M_b^2)\,\frac{m_a\mp  M_a}{m_a \pm M_a} \leq
 s_+ (t_\pm^{(a)}) = (m_a \pm M_a)^2  \,,
 \label{result-spm}
\end{eqnarray}
where the role of $s_+$ and $s_-$ is interchanged in the case that the first inequality in (\ref{result-spm}) is not
realized. Corresponding results for $s_\pm (t_\pm^{(b)}) $ follow from (\ref{result-spm}) by interchanging the indices
$a \leftrightarrow b$.

The analytic continuation is achieved by deforming the $x$-integration contour: a complex contour
$\gamma(z)$ with $\gamma(0)=-1$ and $\gamma(1)=1$ needs to be devised accordingly. At
sufficiently large $s$ the representation (\ref{disp-general-u-t-channel}) is true
always by construction, only as one lowers the energy a deformation of the integration contour is required.
We derive the result
\begin{eqnarray}
&&\int^{1}_{0} d z \,\gamma'(z)\,
\frac{ \lambda_n(s,\,\gamma(z))}{t[s\,,\gamma(z)]\,-m_t^2}=\int_{-1}^{+1} d x \,
\frac{ \lambda_n(s,\,x)}{t_{ab}(s)+2\,x \,p_{a}(s)\,p_{b}(s)}
\nonumber\\
&& \quad \!\!
-i\,\pi\,\frac{ \lambda_n(s,
-\frac{t_{ab}(s)}{2\,p_{a}(s)\,p_{b}(s)} )\,}{p_{a}(s)\,p_{b}(s)}\,
\Theta \Big[\Im \Big(t_{ab}(s)\,p_{a}(s)\,p_{b}(s) \Big)\Big] \,,
\nonumber\\ \nonumber\\
&& t[s\,,x]\,-m_t^2 = t_{ab}(s)+2\,x \,p_{a}(s)\,p_{b}(s)\,,
\label{def-generalized-potential-continued}
\end{eqnarray}
where we consider a typical t-channel process (see also \cite{Karplus:1958zz,Mandelstam:1960zz}) and recall that the deformation of
the $x$-integration contour is required only if the condition (\ref{def-spm}) is realized. The function $t_{ab}(s\,)$ is
defined implicitly in (\ref{def-generalized-potential-continued}).
For $s$ real the function $t_{ab}(s)$ is real
as well. The analytic continuation (\ref{def-generalized-potential-continued})
is valid for $\sqrt{s}> m_a+M_a $ and $\sqrt{s}> m_b+M_b $ at least. For smaller energies the
expressions may demand further modifications. The continuation is necessary since
the function $t_{ab}(s)$ may pass through zero while
$\Im (p_{a}(s)\,p_{b}(s)) \neq 0$. This is the condition (\ref{def-spm}).
If one dropped the second term in
(\ref{def-generalized-potential-continued}) the integral would be discontinuous
right where $t_{ab} (s)= 0$. Such a discontinuity would be
incompatible  with the representation (\ref{disp-general-u-t-channel}).
Note that for the validity of (\ref{def-generalized-potential-continued}) it is irrelevant which of the
various normal or anomalous thresholds in (\ref{central-result}) cause the occurrence of a
zero in $t_{ab}(s)$: in any case the proper result must be continuous at
that zero. A direct  consequence of the analytic continuation is the presence of an anomalous threshold behavior: due to the
second line of (\ref{def-generalized-potential-continued})  the dispersion integral (\ref{disp-general-u-t-channel})
may exhibit a singularity at a threshold or pseudo-threshold energy \cite{Karplus:1958zz,Mandelstam:1960zz}.

It is left to derive the analytic continuation of the r.h.s. (\ref{disp-general-u-t-channel}) mandatory for
(\ref{def-spm}). Using the deformed $x$-integration contour of (\ref{def-generalized-potential-continued}) and
replacing the spectral weight $\varrho^{(t)}_{\pm,i}(m^2,\,m_t^2)$  in (\ref{def-four-cases}) by
\begin{eqnarray}
&& \varrho^{(t)}_{\pm,i}(m^2)\,\Theta[m^2-m_t^2] + \Delta \varrho^{(t-)}_{\pm,i}(m^2,\,m_t^2)+
\Delta \varrho^{(t+)}_{\pm,i}(m^2,\,m_t^2)\,,
\label{def-analytic-continuation-generic-b}
\end{eqnarray}
will ensure the validity of (\ref{disp-general-u-t-channel}) for energies exceeding any of the two normal thresholds.

We consider first the case $t^-_I>0$ with $i=2$ or $i=4$ in (\ref{def-analytic-continuation-generic-b}).
In order to derive the analytic continuation it is useful to establish the inequalities
\begin{eqnarray}
&&t_I^- >  0 \qquad \to \qquad \bar   v_+^+ \geq t_-^+ \geq    v_+^+  \quad {\rm or } \quad \bar v^+_+ < v_+^+  \,,
\nonumber\\ \label{closed-contour-conditions-b}\\
&&  t_-^- \geq t^+_+ \quad \& \quad  v_+^- \geq t_-^- \geq \bar v_+^- \geq {\rm Max} \{v_-^+,\,t^-_+ \} \qquad {\rm if} \quad  \;\; \& \;\; v_-^+ <  v_+^- \,,
\nonumber\\
&&  t_+^+ \geq t^-_- \quad \& \quad v_-^+ \geq t_+^+ \geq \bar v_-^+ \geq {\rm Max} \{v_+^- ,\,t^-_+\} \qquad {\rm if} \quad    \;\;\& \;\; v_-^+ \geq v_+^-\,,
\nonumber
\end{eqnarray}
which suggest the application of the following closed contour
\begin{eqnarray}
&& c^{(t)}_{+,ab}(m^2) \qquad {\rm with} \quad   {\rm Max} \{\bar v_+^+, \,v_+^+\} > m^2 > \left\{
\begin{array}{ll}
v_+^-  \qquad & {\rm if} \qquad v_-^+ <v_+^- \\
v_-^+ \qquad & {\rm if} \qquad v_+^-< v_-^+
\end{array} \right.\,,
\nonumber\\
&& c^{(t)}_{-,ab}(m^2) \qquad {\rm with} \quad   v_+^+ > m^2 >
\left\{
\begin{array}{ll}
\bar v_+^-  \qquad & {\rm if} \qquad v_-^+ <v_+^- \\
\bar v_-^+ \qquad &{\rm if} \qquad v_+^-< v_-^+
\end{array} \right.\,.
\label{def-closed-contour-b}
\end{eqnarray}
First, we assume  $\bar v_+^+ \geq v^+_+ $ for which
the spectral weights $\Delta \varrho^{(t+)}_{\pm,i}(m^2,\,m_t^2)$ are constructed from the condition that the
expression (\ref{def-analytic-continuation-generic-b}) vanishes for $m$ lying on the contours (\ref{def-closed-contour-b}).
We find the result
\begin{eqnarray}
&& \Delta \varrho^{(t+)}_{\pm,2}(m^2,\,m_t^2) =
 \pi\,\frac{\Theta [t^-_- - m_t^2]  }{2\,p_a(s)\,p_b(s)} \,\Bigg|_{s =c^{(t)}_\pm(m^2)}
\nonumber\\
&& \qquad \,\times \Bigg\{ \left( 1- 2\,\Theta[t_I^+-m^2]\,\Theta[t_I^+-v_+^-]  \right)
 \Theta[m^2-v_+^-]\,\Theta[v_+^+-m^2]
\nonumber\\
&& \qquad \;\;\,\, + \left( 2\,\Theta[\pm\,t^\pm_- \mp\,m^2]-1 \right)\Theta[\pm \,\bar v_+^\pm \mp m^2]\,\Theta[\pm \,m^2 \mp v_+^\pm]
\Bigg\}\,,
\nonumber\\
&& \Delta \varrho^{(t+)}_{\pm,4}(m^2,\,m_t^2) = \pi\,\frac{\Theta [t_+^+- m_t^2]}{2\,p_a(s)\,p_b(s)} \,
\Bigg|_{s =c^{(t)}_\pm(m^2)}
\nonumber\\
&& \qquad \,\times \Bigg\{ \left( 2\,\Theta[m^2-t_I^+]\,\Theta[v^+_+-t_I^+]-1 \right)
\Theta[m^2-v_-^+]\,\Theta[v_+^+-m^2]
\nonumber\\
&& \qquad \;\;\,\, + \left( 2\,\Theta[t^+_\mp -m^2]-1 \right)\Theta[\pm \,\bar v_\pm^+ \mp m^2]\,\Theta[\pm \,m^2 \mp v_\pm^+]
\Bigg\}\,.
\label{analytic-continuation-result-b}
\end{eqnarray}
For $\bar v_+^+ < v^+_+ $ the analytic continuation  requires yet the further critical point
\begin{eqnarray}
&& t_0 =  ( M^2_a\,m^2_b - m^2_a\,M^2_b )\,
\frac{m_a^2 - M_a^2 - m_b^2 + M_b^2}{(m_a^2 - M_a^2)\,(m_b^2 - M_b^2)}\,,
\label{def-v0}
\end{eqnarray}
which specifies the exchange mass where either the plus or the minus contour runs through the particular point $s=0$.
The appropriate closed contour defining the desired analytic continuation needs to be extended:
while the minus contour specification in (\ref{def-closed-contour-b}) remains
untouched the plus contour must be enlarged as to include the region $v^+_+ \leq  m^2\leq \infty$. The
results (\ref{analytic-continuation-result-b}) are valid for both cases $\bar v^+_+ \leq v^+_+$ and
$\bar v^+_+ \geq v^+_+$ for masses $m$ in the regions
introduced in (\ref{def-closed-contour-b}). On the extended plus contour the spectral weight is
\begin{eqnarray}
&&\Delta \varrho_{+,2}^{(t+)}(m^2,m_t^2) = \pi\,\frac{\Theta [t_-^-- m_t^2]}{2\,p_a(s)\,p_b(s)} \,
\Bigg|_{s =c^{(t)}_+(m^2)} \left\{
\begin{array}{lll}
+2 \;\; &{\rm for } \;\,&v^+_+ \leq m^2 \leq t_0\\
+1 &{\rm for } &t_0 \leq m^2\leq t^+_- \\
-1 & {\rm for }&t^+_- \leq m^2 \leq \infty
\end{array}
\right. \,,
\nonumber\\ \label{Delta-rho2-4-zero}\\
&& \Delta \varrho_{+,4}^{(t+)}(m^2,m_t^2) = \pi\,\frac{\Theta [t_+^+- m_t^2]}{2\,p_a(s)\,p_b(s)} \,
\Bigg|_{s =c^{(t)}_+(m^2)} \left\{
\begin{array}{lll}
+2\;\; &{\rm for } \;\,&v^+_+ \leq m^2 \leq t_0\\
+1 &{\rm for } &t_0 \leq m^2\leq t^+_- \\
-1 & {\rm for }&t^+_- \leq m^2 \leq \infty
\end{array}
\right. \,, \nonumber
\end{eqnarray}
for $\bar v^+_+\leq v^+_+ \leq  m^2\leq \infty $.

It is left to consider $t_I^-\leq 0$ with $i=1$ or $i=3$ in (\ref{def-analytic-continuation-generic-b}). Four distinct cases arise which are
characterized by the following inequalities
\begin{eqnarray}
&&t_I^- \leq  0 \qquad \to \qquad   \bar   v_+^+ \leq   v_+^+  \qquad  \qquad\; \; {\rm always}\,,
\nonumber\\ \label{def-4-cases}\\
&&  \! \begin{array}{l}
v_+^- \geq t_-^+ \geq \bar v_+^- \geq {\rm Max} \{v_-^+,\,t^-_+ \}  \\
t^-_- \leq t_0 \leq \bar v^+_+ \leq v^-_-
\end{array}
\qquad \quad \;\; {\rm if} \quad  t^+_+ \leq t^+_- \;\; \& \; \;v_-^+ <  v_+^- \,,
\nonumber\\ \nonumber\\
&&  \! \begin{array}{l}
 v_-^+ \geq t_+^+ \geq \bar v_-^+ \geq {\rm Max} \{v_+^-,\,t^-_+\} \\
 t^-_- \leq t_0 \leq \bar v^+_+ \leq v^-_-
\end{array}
 \qquad  \quad \;\;{\rm if} \quad  t^+_+ \geq t^+_-\;\; \& \; \;v_-^+ \geq v_+^-  \,,
\nonumber\\\nonumber\\
&& \! \begin{array}{l}
 v_-^+ \leq t_-^+ \leq \bar v_-^+ \leq v_+^- \quad \& \quad t^-_+ \leq 0  \\
 \bar v_-^- \geq v_-^- \qquad {\rm or} \qquad \bar v_-^- \leq t_-^- \leq  v_-^-
\end{array}
 \qquad {\rm if} \quad  t^+_+ \geq t^+_- \;\; \& \; \;v_-^+ <  v_+^- \,,
\nonumber\\
&& \! \begin{array}{l}
 v_+^- \leq t_+^+ \leq \bar v_+^- \leq v_-^+ \quad \& \quad t^-_+ \leq 0
 \\
 \bar v_-^- \geq v_-^- \qquad {\rm or} \qquad \bar v_-^- \leq t_-^- \leq  v_-^-
\end{array}
 \qquad {\rm if} \quad  t^+_+ \leq t^+_-\;\; \& \; \;v_-^+ \geq v_+^-  \,.
\nonumber
\end{eqnarray}
The first two cases in (\ref{def-4-cases}) lead to a closed contour similar to the one introduced in  (\ref{def-closed-contour-b}). While
the mass range for the minus part in (\ref{def-closed-contour-b}) is unchanged the plus part extends to arbitrarily large and negative
$m^2$. It holds
\begin{eqnarray}
&&  m^2 \leq \bar v^+_+  \qquad {\rm or}\qquad  \,v_+^+ > m^2 > \left\{
\begin{array}{ll}
v_+^-  \qquad & {\rm if} \qquad v_-^+ <v_+^- \\
v_-^+ \qquad & {\rm if} \qquad v_+^-< v_-^+
\end{array} \right.\,.
\label{analytic-continuation-result-b-modified}
\end{eqnarray}
The last two cases in (\ref{def-4-cases}) involve yet the further closed contour
\begin{eqnarray}
&& c^{(t)}_{-,ab}(m^2) \qquad {\rm with} \quad   v_-^- < m^2 <
\left\{
\begin{array}{ll}
\bar v_-^+  \qquad & {\rm if} \qquad v_-^+ <v_+^- \\
\bar v_+^- \qquad &{\rm if} \qquad v_+^-< v_-^+
\end{array} \right. \,,
\label{def-closed-contour-c}\\
&& c^{(t)}_{+,ab}(m^2) \qquad {\rm with} \quad   {\rm Min} \{\bar v_-^-,\, v^-_-\} < m^2 < \left\{
\begin{array}{ll}
v_-^+  \qquad & {\rm if} \qquad v_-^+ <v_+^- \\
v_+^- \qquad & {\rm if} \qquad v_+^-< v_-^+
\end{array} \right.\,.
\nonumber
\end{eqnarray}
For $\bar v^-_- \leq v^-_-$ the spectral weights $\Delta \varrho^{(t+)}_{\pm,i}(m^2,\,m_t^2)$  with $i=1$ and $i=3$ follow from the
requirement  that the expressions (\ref{def-analytic-continuation-generic-b})  vanish for exchange masses $m$ on the closed contour as
specified in (\ref{def-closed-contour-b}) or (\ref{def-closed-contour-c}) depending on the specifics of the case. Some algebra leads to
\begin{eqnarray}
&& \Delta \varrho^{(t+)}_{\pm,1}(m^2,\,m_t^2) =
\pi\,\frac{\Theta [\,t^+_-- m_t^2] }{2\,p_a(s)\,p_b(s)} \,\Bigg|_{s =c^{(t)}_\pm(m^2)}
\, \Theta [ \,t^+_- -t^+_+ ]
\nonumber\\
&& \qquad \,\times  \Bigg\{ - \Theta[m^2-v_+^-]\,\Theta[v_+^+-m^2] + \Big(2\,\Theta[\mp \,m^2 \pm  t^\mp_- ] -1
\nonumber\\
&&  \qquad \qquad \qquad \;\;-\, \Theta[ \pm\,m^2 \mp t_0 ] \Big)\,
 \Theta[\pm \,\bar v_+^\pm \mp m^2]\,\Theta[v_+^\pm - m^2] \Bigg\}
\nonumber\\
&& \qquad \qquad \qquad \;\;+ \,\pi\,\frac{\Theta [\,t^-_- - m_t^2]  }{2\,p_a(s)\,p_b(s)} \,\Bigg|_{s =c^{(t)}_\pm(m^2)}
\Theta [t^+_+ - t^+_-]
\nonumber\\
&& \qquad \,\times \Bigg\{ \left( 2\,\Theta[m^2- t_I^+]\,\Theta[v^+_--t_I^+]-1 \right)
 \Theta[m^2-v_-^-]\,\Theta[v^+_- -m^2]
\nonumber\\
&& \qquad \qquad \qquad \;\; + \left( 2\,\Theta[\pm\,t^\mp_- \mp\,m^2]-1 \right) \Theta[\pm \,m^2 \mp \bar v_-^\mp]\,\Theta[\pm \,v_-^\mp \mp m^2]
\Bigg\}\,,
\nonumber\\
&& \Delta \varrho^{(t+)}_{\pm,3}(m^2,\,m_t^2)= \pi\,\frac{ \Theta [\,t_+^+- m_t^2]}{2\,p_a(s)\,p_b(s)} \,
\Bigg|_{s =c^{(t)}_\pm(m^2)} \,\Theta [\,t^+_+ - t^+_-]
\nonumber\\
&& \qquad \,\times  \Bigg\{- \Theta[m^2-v_-^+]\,\Theta[v_+^+-m^2] + \Big(2\,\Theta[\,t^\mp_\mp - m^2  ] -1
\nonumber\\
&& \qquad \qquad \qquad \;\; - \,\Theta[ \pm \,m^2 \mp \,t_0 ]\Big)\, \Theta[\pm \,\bar  v_\pm^+ \mp m^2]\,\Theta[v_\pm^+ - m^2]
\Bigg\}
\nonumber\\
&& \qquad \qquad \qquad \;\;+\,\pi\,\frac{\Theta [\,t_-^-- m_t^2]}{2\,p_a(s)\,p_b(s)} \,
\Bigg|_{s =c^{(t)}_\pm(m^2)} \,\Theta [\, t^+_- -t^+_+ ]
\nonumber\\
&& \qquad \,\times \Bigg\{ \left( 2\,\Theta[\,m^2-t_I^+] -1 \right)
\Theta[m^2-v_-^-]\,\Theta[v^-_+ -m^2]
\nonumber\\
&& \qquad \qquad \qquad \;\; + \left( 2\,\Theta[t^\mp_\mp -m^2]-1 \right)\Theta[\pm \,m^2 \mp \bar v^-_\mp]\,\Theta[\pm \,v^-_\mp \mp m^2]
\Bigg\}\,.
\label{analytic-continuation-result-c}
\end{eqnarray}
The results (\ref{analytic-continuation-result-c}) are valid for both cases $\bar v^-_- \leq v^-_-$ and
$\bar v^-_- \geq v^-_-$ for masses $m$ in the regions
introduced in (\ref{def-closed-contour-b}) and (\ref{def-closed-contour-c}). A generalization is needed if the
contour (\ref{def-closed-contour-c}) is probed with $\bar v^-_- \geq v^-_-$. While the minus contour region is
unchanged the plus contour of (\ref{def-closed-contour-c}) is modified to cover the additional interval
$t^-_- \leq m^2 \leq v^-_-$.  On the extended plus contour the spectral weight is
\begin{eqnarray}
&&\Delta \varrho_{+,1}^{(t+)}(m^2,m_t^2) = \pi\,\frac{\Theta [\,t_+^+- t^+_-]}{2\,p_a(s)\,p_b(s)} \,
\Bigg|_{s =c^{(t)}_+(m^2)} \left\{
\begin{array}{lll}
- 2 \;\; &{\rm for } \;\,& t^-_- \leq m^2 \leq v^-_-\\
0 &{\rm for } &  m^2\leq t^-_-
\end{array}
\right. \,,
\nonumber\\ \label{Delta-rho1-3-zero}\\
&& \Delta \varrho_{+,3}^{(t+)}(m^2,m_t^2) = \pi\,\frac{\Theta [ \,t^+_- -t_+^+]}{2\,p_a(s)\,p_b(s)} \,
\Bigg|_{s =c^{(t)}_+(m^2)} \left\{
\begin{array}{lll}
- 2\;\; &{\rm for } \;\,&t^-_- \leq m^2 \leq v^-_-\\
0  &{\rm for } & m^2\leq t^-_-
\end{array}
\right. \,, \nonumber
\end{eqnarray}
for $ m^2 \leq v^-_- $ and $m_t^2 \leq  t^-_- $.

\clearpage

\section{Spectral representation:  some examples }

\begin{figure*}[b]
\center{
\includegraphics[keepaspectratio,width=0.9\textwidth]{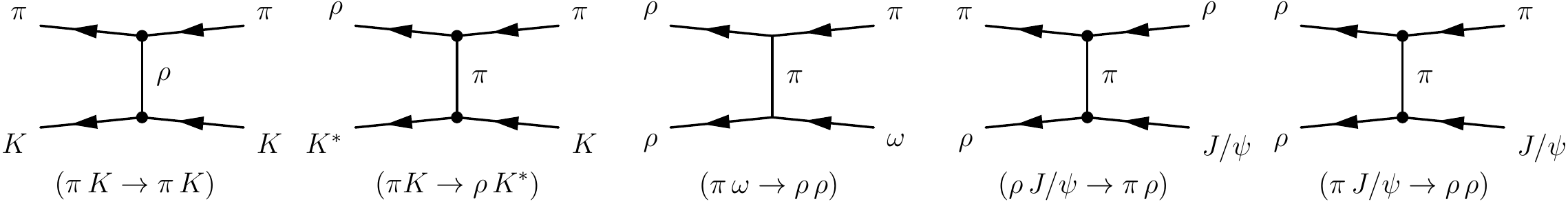} }
\caption{\label{fig:diag-list-t} Some specific t-channel exchange processes.  }
\end{figure*}

\begin{figure*}[b]
\center{
\includegraphics[keepaspectratio,width=0.9\textwidth]{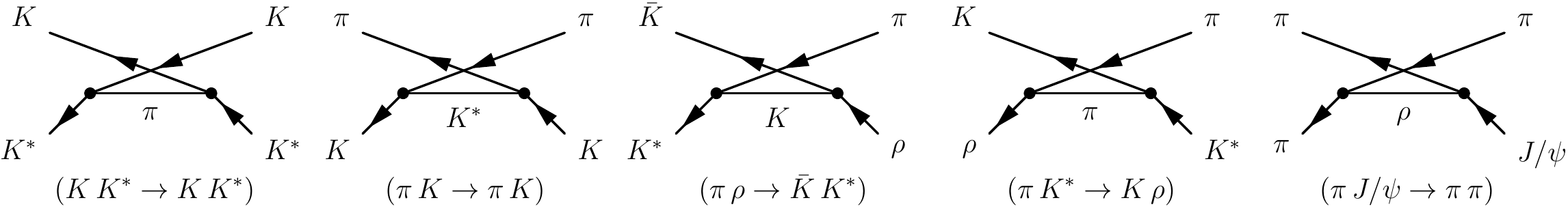} }
\caption{\label{fig:diag-list-u} Some specific u-channel exchange processes.  }
\end{figure*}

In this section we illustrate the formalism developed above at hand of several selected reactions with specific
contributions from $t$- and u-channel exchange processes. We pick reactions which give a good
illustration of the various cases summarized in the general spectral density (\ref{central-result}, \ref{def-analytic-continuation-generic-b}).
In Fig. \ref{fig:diag-list-t} and Fig. \ref{fig:diag-list-u} our choices are shown. We consider five t-channel and five u-channel processes
involving pseudo-scalar and vector particles.

In a first step we compute the list of critical exchange masses and collect them in Tab. \ref{tab:critical-points-t}  for the t-channel 
processes. For later convenience the critical points are labeled from 1 to 15. Isospin averaged particle masses from the PDG are
used. All dimension full quantities are expressed in units of the isospin averaged pion masses. A critical exchange mass is not always
active in the expression (\ref{central-result}). Only if it is larger than the mass of the
exchange particle it may turn relevant. The extra terms introduced in (\ref{def-analytic-continuation-generic-b}) 
imply a further analytic continuation of the dispersion integral (\ref{disp-general-u-t-channel}). While the additional terms 
(\ref{analytic-continuation-result-a}) are required for the validity of (\ref{disp-general-u-t-channel})  
slightly above the largest threshold point, the extra contributions 
(\ref{analytic-continuation-result-b}, \ref{analytic-continuation-result-c}) are necessary to realize  
(\ref{disp-general-u-t-channel}) in-between the two nominal thresholds. In the absence of such terms 
(\ref{disp-general-u-t-channel}) holds only for large $s$ exceeding some critical value. 
All extra terms (\ref{analytic-continuation-result-a}, \ref{analytic-continuation-result-b}, \ref{analytic-continuation-result-c}) 
will be illustrated by our t-channel examples. 

\def\phx{\phantom{xx}}
\def\phm{\phantom{$-$}}
\begin{table}[t]
\begin{tabular}{l|cccccc}
\hline\hline
 & &\phx$\pi \,K\to \pi\, K$\phx & \phx $\pi \,K\to\rho\, K^*$\phx &\phx$\pi\, \om\to\rho \,\rho$\phx & 
 \phx $\rho\, J/\psi\to\pi\,\rho$\phx  &\phx $\pi\, J/\psi\to\rho\,\rho$\\
 \hline\hline
1  & $t^+_+$ & 0 & \phm29.8812 & \phm25.4465 & \phm97.3278 & \phm221.066 \\
2  & $t^+_-$ & 0 & \phm30.5313 & $\infty$    & \phm116.185 & $\infty$  \\
3  & $t^-_+$ & 0 & $-$29.6174  & $-$14.5458  & $-$118.153  & \phm8.67998 \\
4  & $t^-_-$ & 0 & $-$36.8085  & \phm36.7849 & $-$66.4585  & \phm53.5515 \\
5  & $t^+_I$ & 0 & \phm30.2063 & $\infty$    & \phm15.4347 & $\infty$  \\
6  & $t^-_I$ & 0 & $-$33.2129  & \phm31.1157 & $-$0.984286 & \phm31.1157 \\
\hline
7  & $v^+_+$ & 51.5765 & \phm101.333 & \phm126.503 & \phm784.778 & \phm784.778 \\
8  & $v^+_-$ & 0       & \phm20.9594 & \phm20.9594 & \phm284.179 & \phm284.179 \\
9  & $v^-_+$ & 4       & \phm43.2720 & \phm43.2720 & \phm43.2720 & \phm43.2720 \\
10 & $v^-_-$ & 0       & \phm8.32172 & \phm0.00829 & \phm20.9594 & \phm20.9594 \\
\hline
11 & $\bar{v}^+_+$ & 0             & \phm8.29860 & $-$15.6877  & \phm20.3388 & $-$118.153 \\
12 & $\bar{v}^+_-$ & Indeterminate & \phm43.2466 & \phm36.4702 & \phm43.8217 & \phm116.185 \\
13 & $\bar{v}^-_+$ & 0             & \phm20.9722 & \phm25.6933 & \phm279.869 & \phm97.3278 \\
14 & $\bar{v}^-_-$ & Indeterminate & \phm101.577 & $-$1812.17  & \phm826.092 & $-$66.4585 \\
\hline
15\; & $t_0$ & 0 & $-$3.00669 & $\infty$  & \phm14.4504 & $\infty$ \\
 \hline\hline
\end{tabular}
\caption{Critical points for the $t$ channel exchange processes shown in Fig.~\ref{fig:diag-list-t} in units $m_\pi^2$}
\label{tab:critical-points-t}
\end{table}

\begin{figure*}[t]
\includegraphics[height=5.7cm]{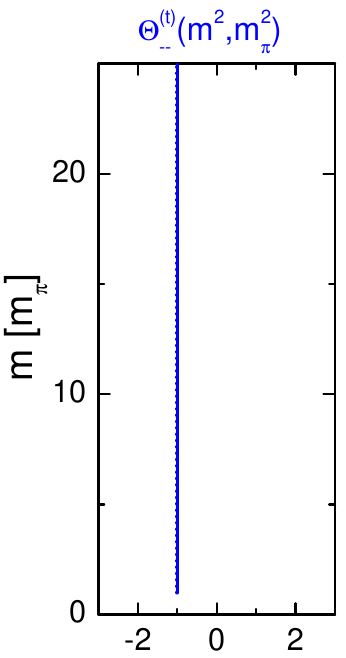}
\includegraphics[height=5.7cm]{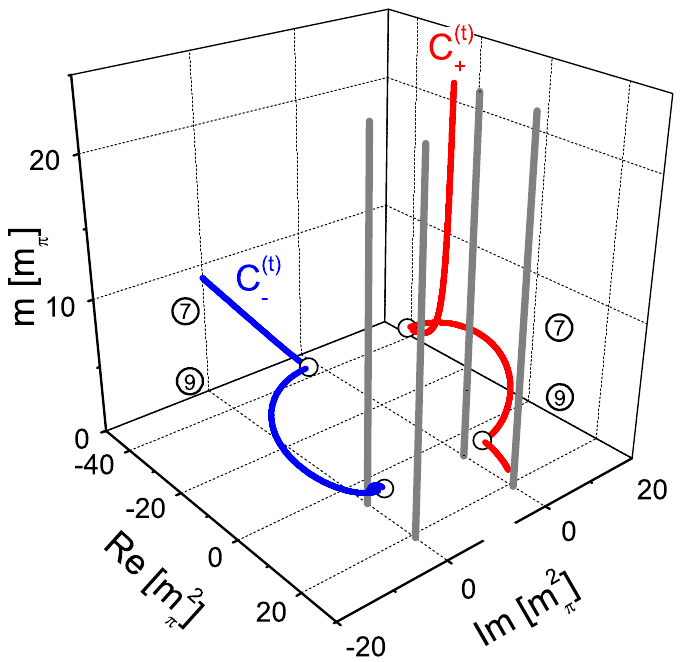}
\includegraphics[height=5.7cm]{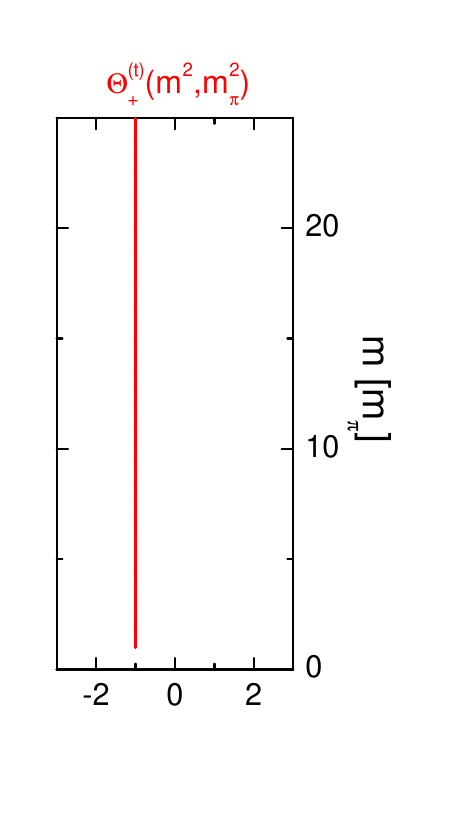}\\ \vskip1cm
\includegraphics[height=5.7cm]{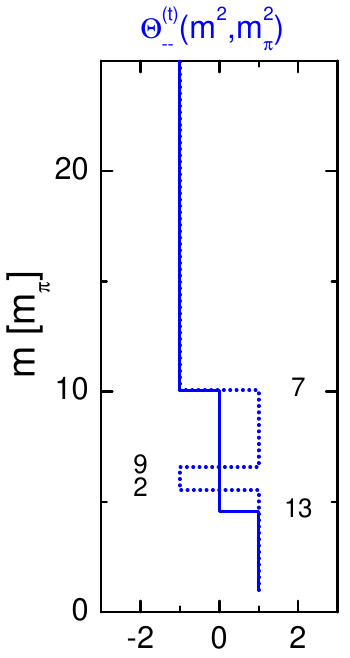}
\includegraphics[height=5.7cm]{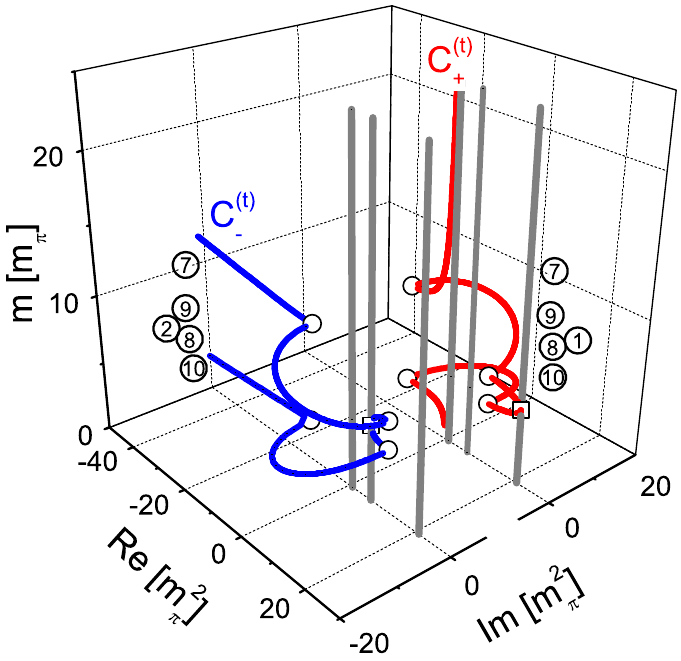}
\includegraphics[height=5.7cm]{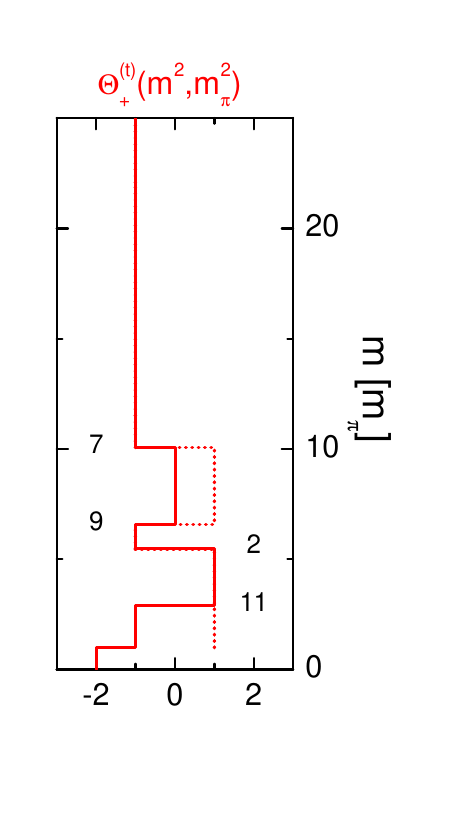} 
\caption{Spectral signatures $\Theta^{(t)}_\pm (m^2, m_\pi^2)$ of (\ref{central-result}, \ref{def-Theta-signature-t}) along the $c_-^{(t)}(m^2)$ (left column) and $c_+^{(t)}(m^2)$ (right column) contours for the t-channel 
processes $\pi\,K \rightarrow \pi\, K$  (upper panel) and $\pi\,K \rightarrow \rho\,K^*$ (lower panel) as functions of the mass of the exchanged particle $m$. The form of the two contours 
are shown in the center of the figure always. The thin pillars show the positions of relevant thresholds or pseudo-thresholds. }
\label{fig:t-channel:A}
\end{figure*}

We briefly discuss the t-channel processes characterized by the list of critical exchange points Tab. \ref{tab:critical-points-t}.
For our first t-channel reaction $\pi\, K \to \pi\, K$ two
critical points $v^+_+$ and $v^-_+$, which are number 7 and 9, may be relevant for both contour lines $c ^{(t)}_+(m^2)$ and $c ^{(t)}_-(m^2)$.
Since the square of the exchange mass, the $\rho$-meson mass, is larger than $v^-_+$ there is only $v^+_+$ left. For this example  it holds
$t^-_I\leq 0 $ and $v^+_- < v^-_+$ so that the corresponding spectral weight is given by the first 
case in (\ref{central-result}). The contour lines  are shown in the center of 
Fig. \ref{fig:t-channel:A}. The two contours leave the real axis at $m^2 =v^-_+$ . At the second critical point $m^2 = v^+_+$ the two 
contours return to the real axis. As  a consequence the spectral weight develops an imaginary part for $v^-_+ \leq m^2 \leq v^+_+$  
only. In order to illustrate the characteristics of the spectral weights $\varrho^{(t)}_\pm (m^2, m_t^2)$ we introduce their signature with
\begin{eqnarray}
\Theta^{(t)}_\pm (m^2, m_t^2) = 2\,\frac{ p_a(s)\,p_b(s)}{\pi}\,\varrho^{(t)}_\pm (m^2, m_t^2)\, \Bigg|_{s = c^{(t)}_\pm (m^2)}\,,
\label{def-Theta-signature-t}
\end{eqnarray}
which is an integer number depending on $m^2$ and $m^2_t$. For our first t-channel example both signatures are set to $-1$ and 
remain unchanged throughout the contours.

For our second t-channel process $\pi\, K \to \rho\, K^*$ there are 6 relevant critical points in the contour paths. The latter are indicated  
in Fig. \ref{fig:t-channel:A} by their label number 1, 2, 7, 8, 9, 10. The positions of the circled numbers in the plot correspond to their 
numerical values as given in Tab. \ref{tab:critical-points-t}. All such points are larger than the square of the exchange mass, in 
this case the pion mass, and therefore a proper evaluation of the spectral weight depends on those  6 critical points. 
The contour paths are off the real axis within the two intervals $ v^-_- < m^2 < v^+_- $ and $ v^-_+ < m^2 < v^+_+ $ only. In general, 
any of the critical points 7, 8, 9, 10 signals that the contour leaves or returns to the real axis. The corresponding critical contour points 
are surrounded by open circles in our plots. The anomalous point 2 characterizes the exchange mass $m^2= t^+_-$  at which the $-$ contour hits a 
threshold or pseudo-threshold pillar. Similarly, the condition $m^2= t^+_+$ specifies the anomalous point 1 at which the $+$ contour hits a 
threshold or pseudo-threshold pillar. In general either of the two contours may touch a threshold or pseudo-threshold pillar only at any of 
the anomalous points 1, 2, 3, 4. If this occurs the contour runs towards the threshold along the real axis till it hits it and then inverts 
the direction and runs away from the threshold again. Whenever this happens the corresponding critical contour point is surrounded by an 
open square in our plots. 

The associated spectral signatures are shown again left and right of the contour paths in the center of  Fig. \ref{fig:t-channel:A}. 
Like for our first example we have $t^-_I \leq 0 $ and $v^+_- < v^-_+$ and the corresponding spectral weights are given by the case 1  
in (\ref{central-result}, \ref{def-analytic-continuation-generic-b}). In contrast, however, the spectral signatures change several times now. 
In our plots the corresponding critical points are indicated by their label number as introduced in Tab. \ref{tab:critical-points-t}. 
In particular, the extra terms (\ref{analytic-continuation-result-c}) prove relevant here. This is so since $t^+_- > m_\pi^2$. We identify 
the associated closed contour. According to (\ref{def-closed-contour-b}) the  $-$ part is specified by $ \bar v^-_+< m^2 < v^+_+$. 
The $+$ part receives two distinct contributions with $ v^-_+< m^2 < v^+_+$ and $ m^2 < \bar v^+_+$ (see (\ref{analytic-continuation-result-b-modified})). 
As a consequence of $t^+_- > t^+_+$ the additional critical points $t_0$ and $\bar v^\pm_+$ are activated in (\ref{analytic-continuation-result-c}). 
Indeed at $m^2 =\bar v^-_+$ and $m^2 =\bar v^+_+$ the $-$ and $+$  
spectral signatures are discontinuous respectively. While in Fig. \ref{fig:t-channel:A} the full lines show the full spectral signatures in the 
presence of the extra terms, the dotted lines show the results implied by (\ref{central-result}) only. We do not show possible contributions 
for $m^2 < 0$ in the plots for the clarity of the presentation. Due to the condition $ m^2 < \bar v^+_+$ discussed above they are present 
nevertheless in the + spectral signature.

\begin{figure*}[t]
\includegraphics[height=5.7cm]{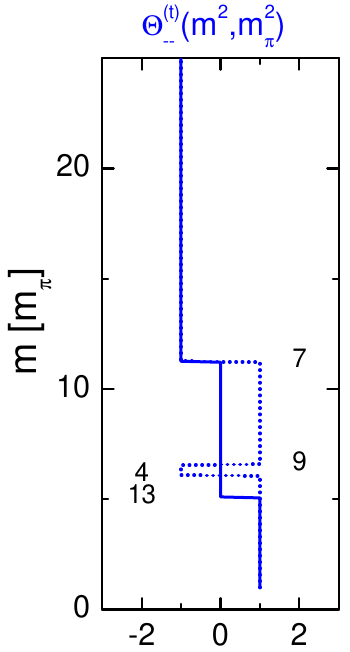}
\includegraphics[height=5.7cm]{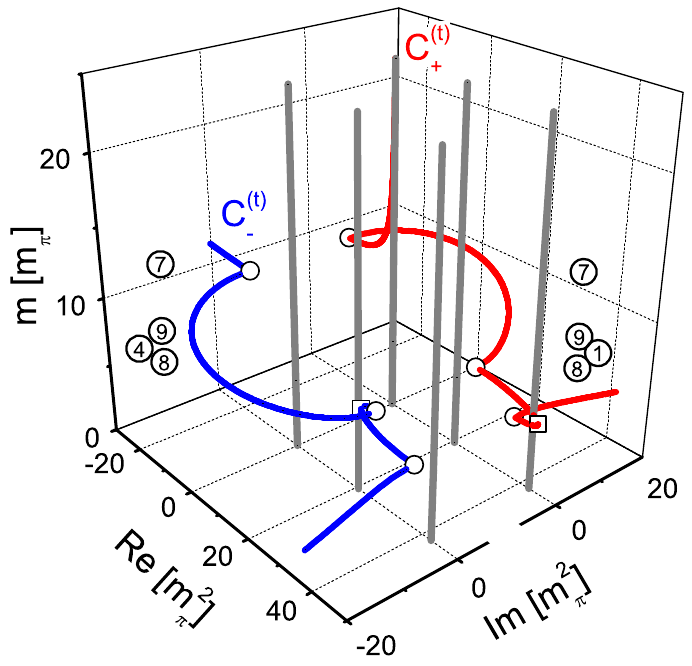}
\includegraphics[height=5.7cm]{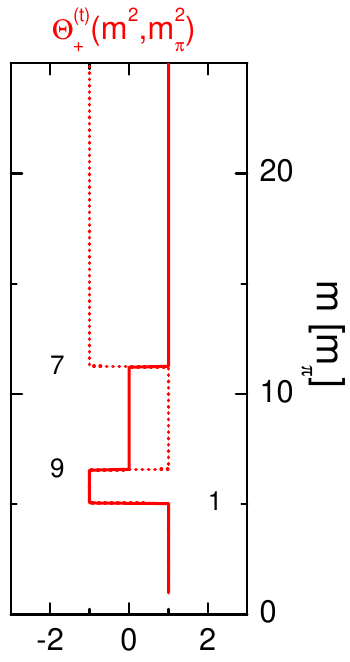} \\ \vskip1cm
\includegraphics[height=5.7cm]{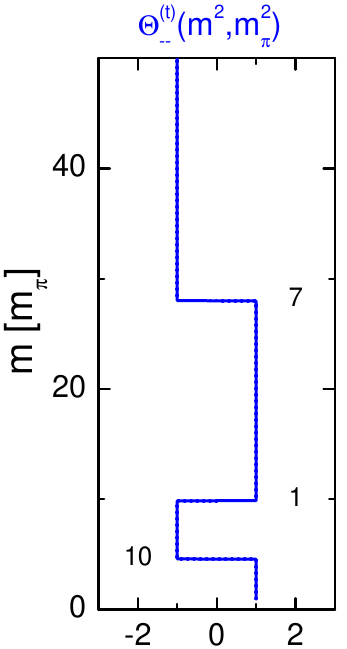}
\includegraphics[height=5.7cm]{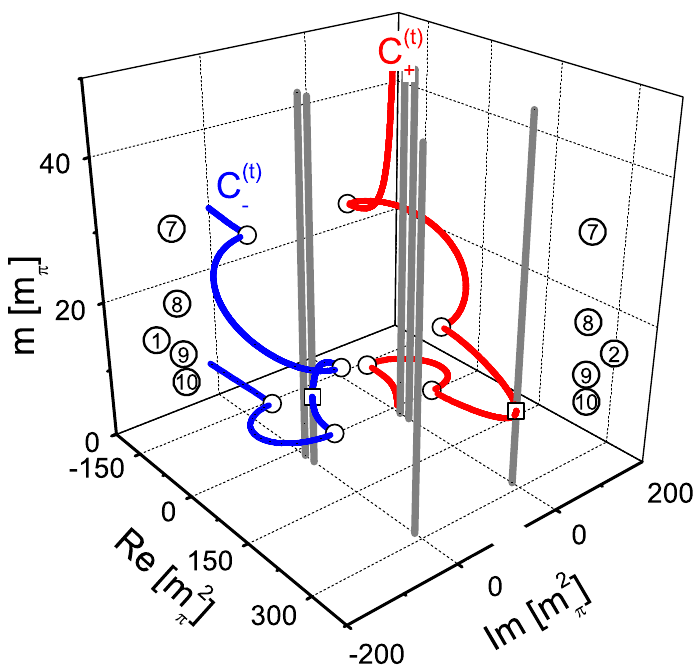}
\includegraphics[height=5.7cm]{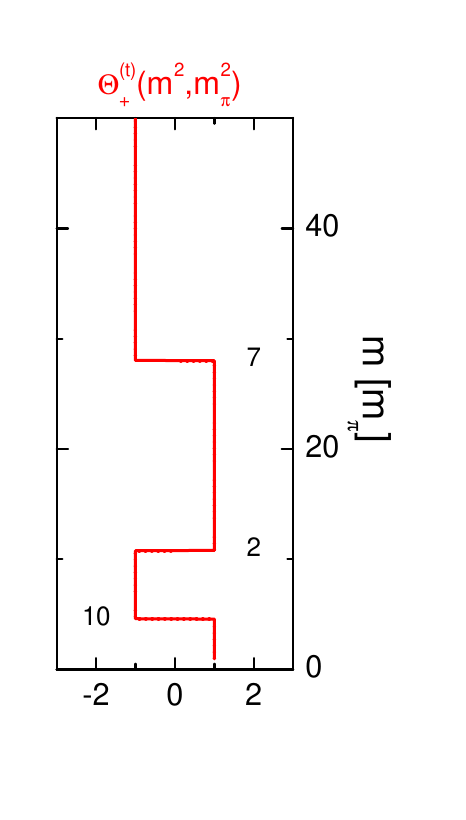} \\ \vskip1cm
\includegraphics[height=5.7cm]{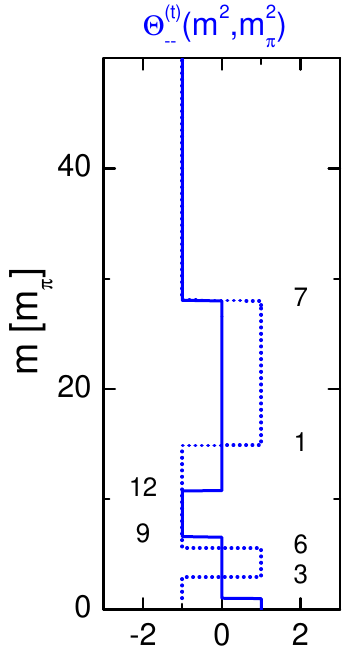}
\includegraphics[height=5.7cm]{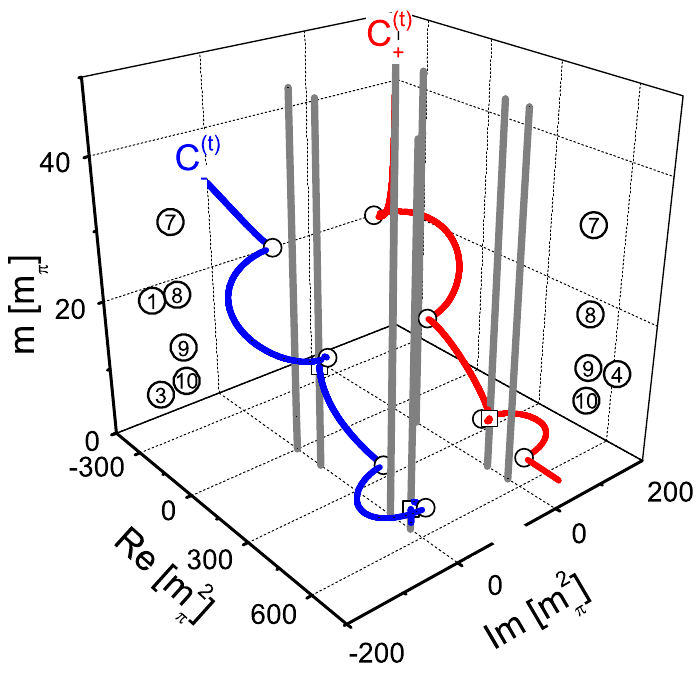}
\includegraphics[height=5.7cm]{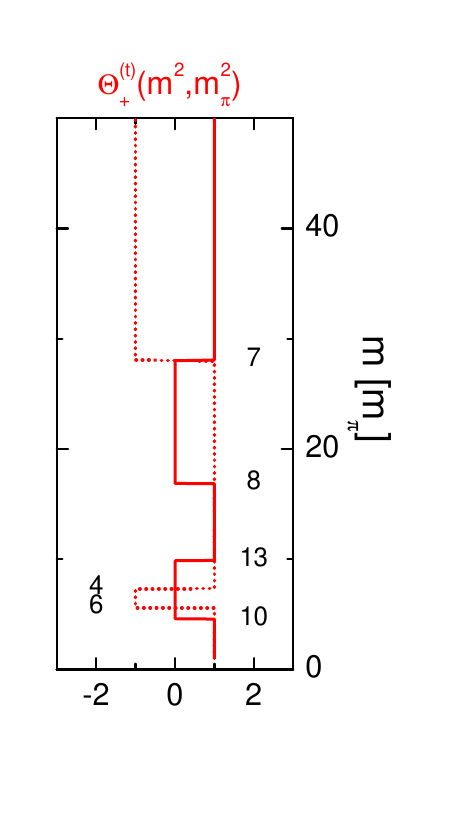} 
\caption{Spectral signatures $\Theta^{(t)}_\pm (m^2, m_\pi^2)$ of (\ref{central-result}, \ref{def-Theta-signature-t}) along the $c_-^{(t)}(m^2)$ (left column) and $c_+^{(t)}(m^2)$ (right column) contours for the t-channel processes 
$\pi \,\omega \rightarrow \rho\,\rho$, $\rho \,J/\Psi \rightarrow \pi\, \rho$ and
$\pi \,J/\Psi \rightarrow \rho\,\rho$ respectively. The form of the two contours 
are shown in the center of the figure always. The thin pillars show the positions of relevant thresholds or pseudo-thresholds.}
\label{fig:t-channel:B}
\end{figure*}

We turn to the remaining three t-channel processes that are illustrated in Fig. \ref{fig:t-channel:B}.  
The corresponding spectral signatures are shown left and right of the contour lines in the center of  the plots. 
In all cases there are non-trivial changes of the signatures at the various critical points. 
The t-channel process  $\pi \,\omega \to \rho \,\rho$ is characterized by the 
critical contour points 1, 4, 7, 8, 9 and the condition $t^-_I > 0$. 
With $v^+_- < v^-_+ < v^+_+$ case 2 in (\ref{central-result}) is implied.
The conditions $m^2 = t^+_+$ and $m^2 = t^-_-$ identify at which point the $+$ and $-$ contours hit a 
threshold pillar respectively. Moreover, the extra terms (\ref{analytic-continuation-result-b}) are active since 
it holds $t^-_- > m_\pi^2$. The corresponding closed contour is identified in (\ref{def-closed-contour-b}, \ref{Delta-rho2-4-zero}) which leads to 
$ \bar v^-_+ < m^2< v^+_+ $ and $ v^-_+ < m^2 $ for the $-$ and $+$ parts respectively. 

With the t-channel process $\rho \,J/\Psi \to \pi \,\rho$ we have an example for case 3 in (\ref{central-result}). The 6 
critical contour points 1, 2, 7, 8, 9, 10 are active and we have $t^-_I \leq 0$ together with $v^+_- > v^-_+$. 
The extra terms in (\ref{analytic-continuation-result-c}) are not probed here since we have $t^+_+ < t^+_-$ together with 
$t^-_- < m_\pi^2$.  
Finally the t-channel process $\pi\, J/\Psi \to \rho\, \rho$ illustrates case 4 in (\ref{central-result}). The 7 
critical contour points 1, 3, 4, 7, 8, 9, 10 are active and we have $t^-_I > 0$ with $v^+_- > v^-_+$. 
Since $t^+_+ >t^-_+ > m_\pi^2$ the extra terms in (\ref{analytic-continuation-result-a}) and in (\ref{analytic-continuation-result-b}) are probed.  
While the first closed contour is specified with  $ \bar v^-_- < m^2 < v^-_+$ and $v^-_- < m^2 < \bar v⁻_+$,
the second closed contour is associated with the conditions $\bar v^+_- < m^2 < v^+_+$ and  $v^+_- < m^2  $ for the 
$-$ and $+$ parts respectively.

\clearpage

\begin{figure*}[t]
\includegraphics[height=5.7cm]{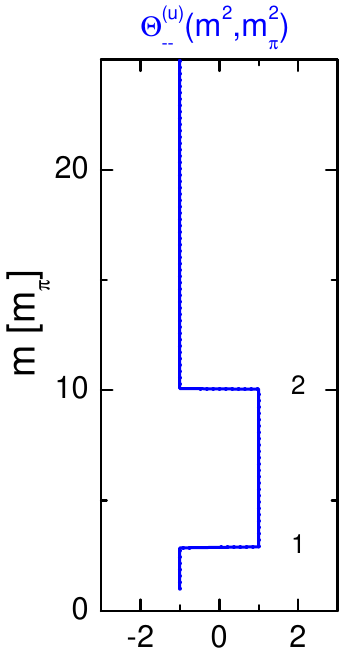}
\includegraphics[height=5.7cm]{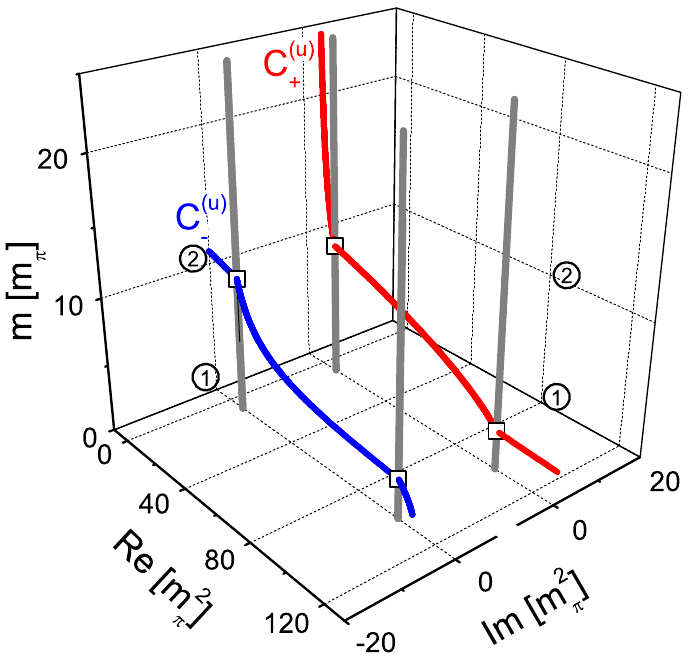}
\includegraphics[height=5.7cm]{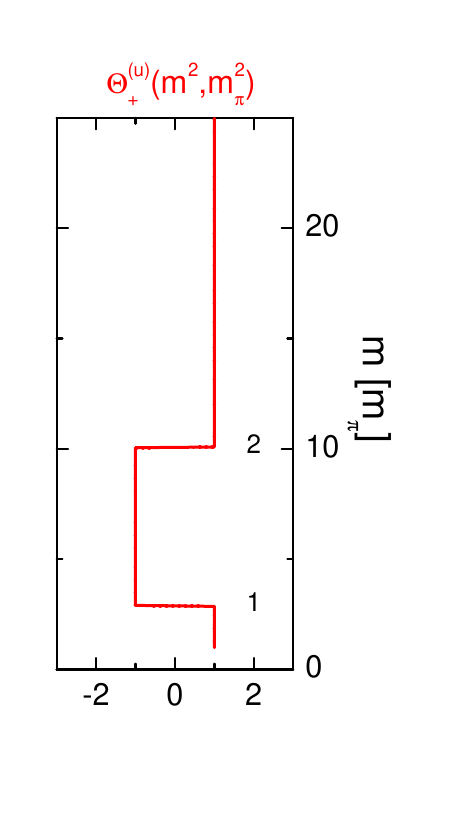} \\ \vskip1cm 
\includegraphics[height=5.7cm]{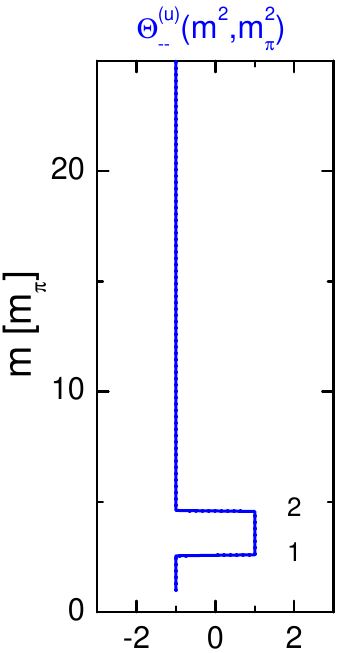}
\includegraphics[height=5.7cm]{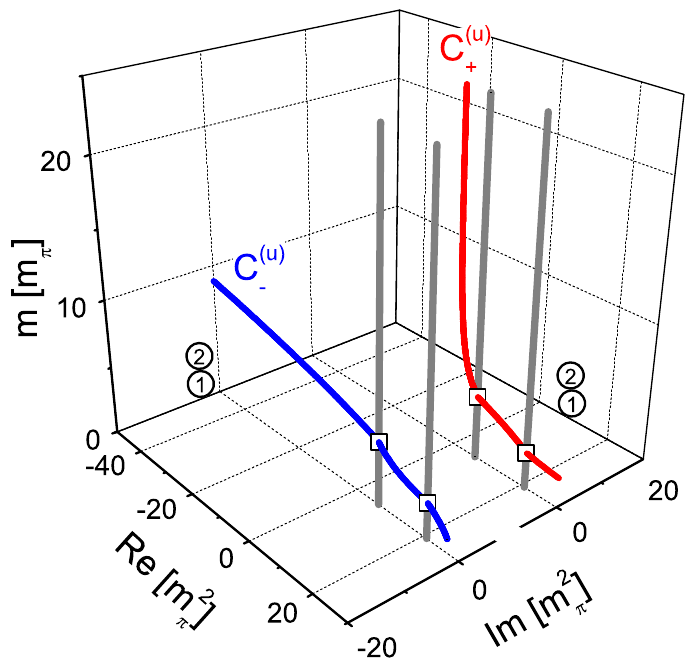}
\includegraphics[height=5.7cm]{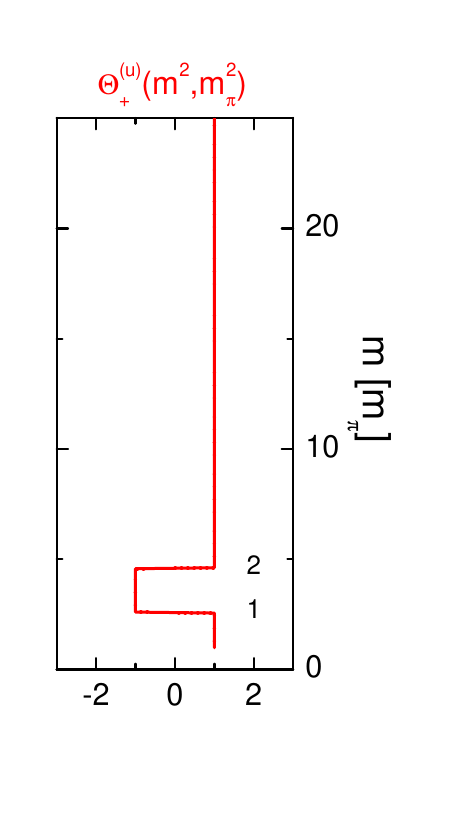}
\caption{Spectral signatures $\Theta^{(u)}_\pm (m^2, m_\pi^2)$  along the $c_-^{(u)}(m^2)$ (left column) and $c_+^{(u)}(m^2)$ (right column) 
contours for the u-channel
processes   $K\, K^* \rightarrow K \,K^*$ and $\pi\,K \rightarrow \pi\,K$ as functions of the mass of the exchanged particle $m$. The form of the two contours 
are shown in the center of the figure always. The thin pillars show the positions of relevant thresholds or pseudo-thresholds.}
\label{fig:u-channel:A}
\end{figure*}

\begin{figure*}[t]
\includegraphics[height=5.7cm]{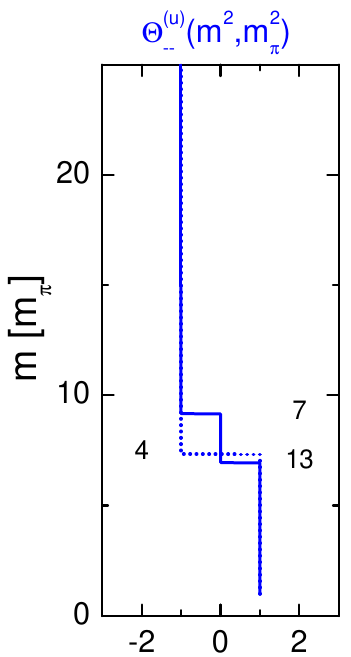}
\includegraphics[height=5.7cm]{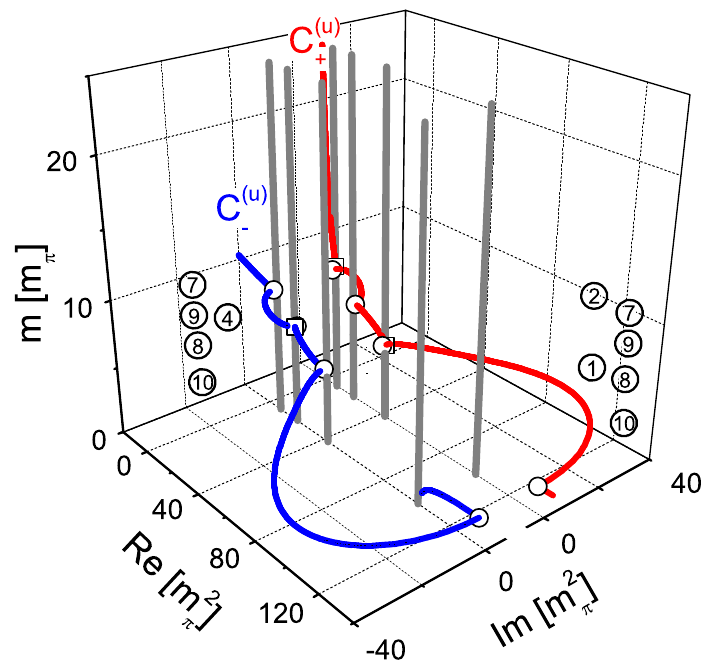}
\includegraphics[height=5.7cm]{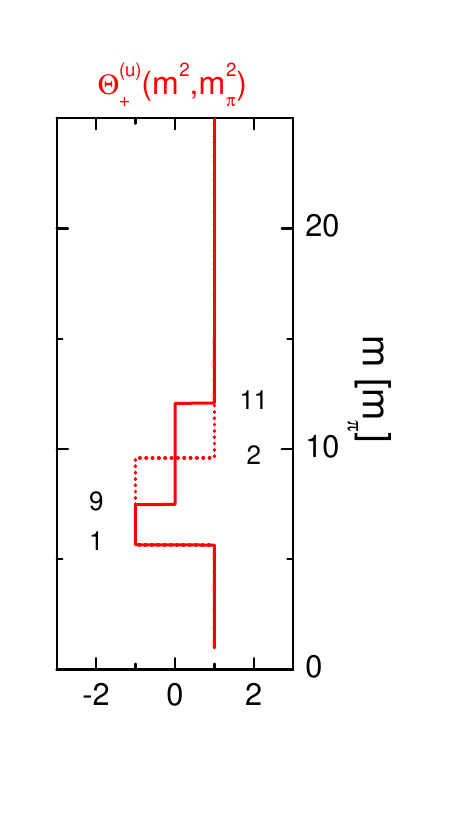} \\ \vskip1cm
\includegraphics[height=5.7cm]{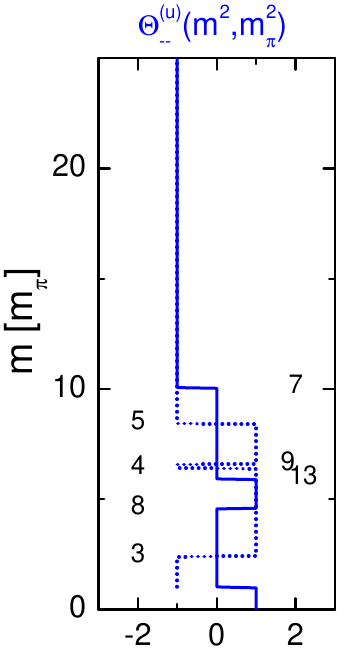}
\includegraphics[height=5.7cm]{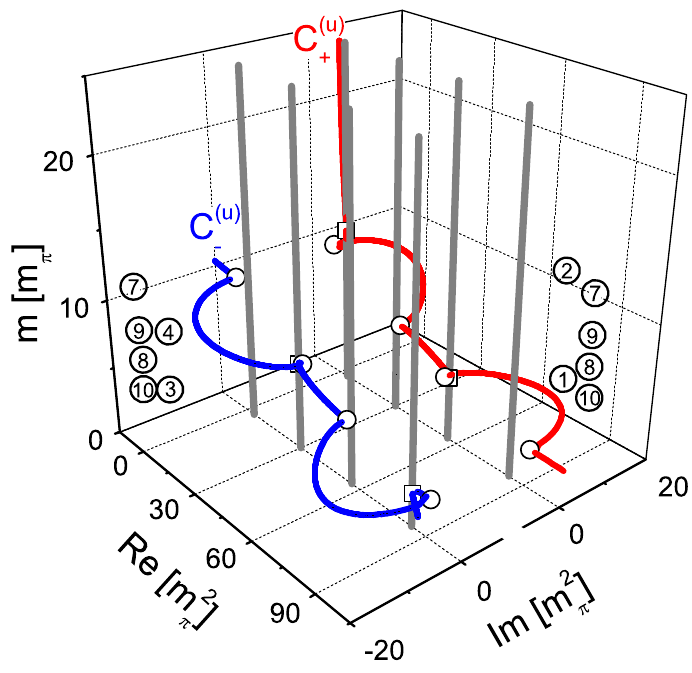}
\includegraphics[height=5.7cm]{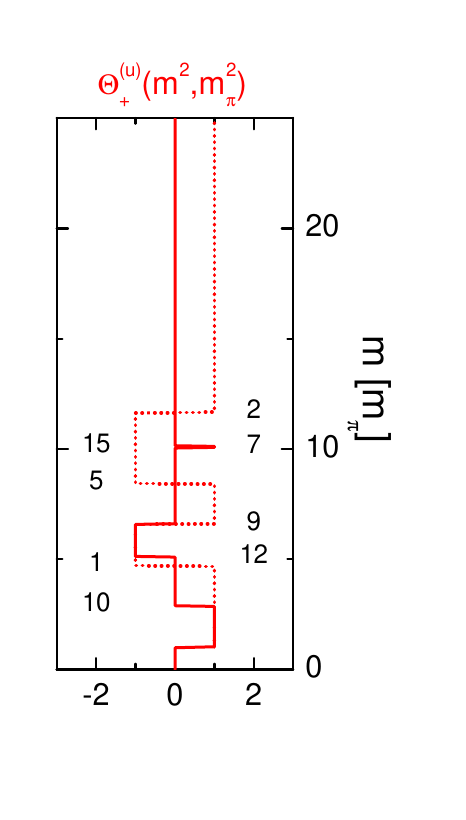} \\ \vskip1cm
\includegraphics[height=5.7cm]{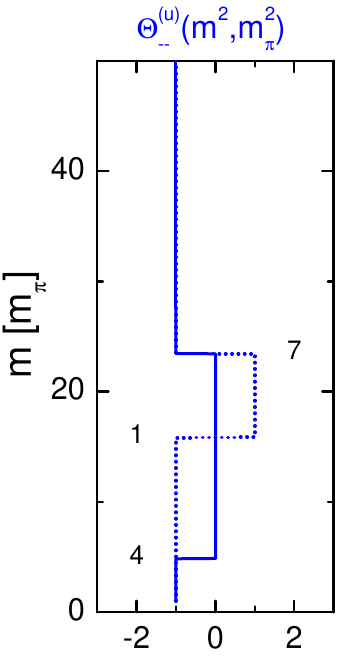}
\includegraphics[height=5.7cm]{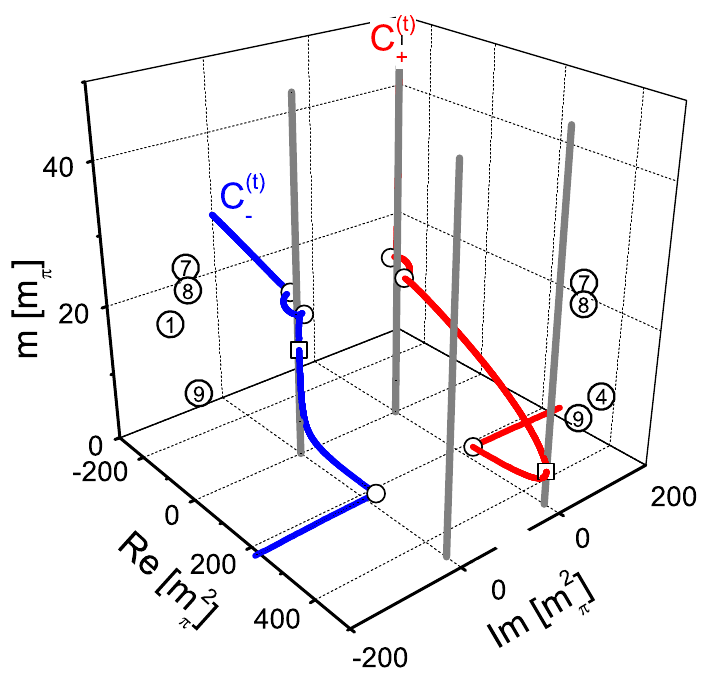}
\includegraphics[height=5.7cm]{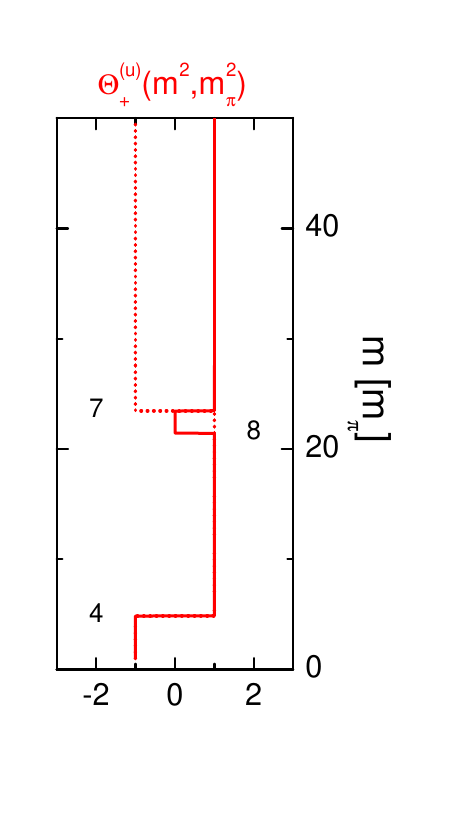} 
\caption{Spectral signatures $\Theta^{(u)}_\pm (m^2, m_\pi^2)$  along the $c_-^{(u)}(m^2)$ (left column) and $c_+^{(u)}(m^2)$ (right column) contours 
for the u-channel processes $\pi\rho \rightarrow \bar K K^*$, $\pi K^* \rightarrow K\rho$ and $\pi J/\Psi \rightarrow \pi\pi$. The form of the two contours 
are shown in the center of the figure always. The thin pillars show the positions of relevant thresholds or pseudo-thresholds.}
\label{fig:u-channel:B}
\end{figure*}

We turn to the u-channel processes of Fig. \ref{fig:diag-list-u}. 
The list of critical exchange masses is collected in Tab. \ref{tab:critical-points-u},  where again the critical points are 
labeled through from 1 to 15. All dimension full quantities are expressed in units of the isospin averaged pion masses. 
We recall that a critical exchange mass is not necessarily active in the expression (\ref{central-result}). 
Only if it is larger than the mass of the exchange particle it may turn relevant. 
The additional terms in the spectral density (\ref{def-analytic-continuation-generic-b}) as constructed 
in  (\ref{analytic-continuation-result-a}, \ref{analytic-continuation-result-b}, \ref{analytic-continuation-result-c}) 
will be needed for our example cases. Our results are illustrated with Fig. \ref{fig:u-channel:A} and 
\ref{fig:u-channel:B} where besides the contour paths in the center of the plots the signatures of the spectral weights 
as introduced with
\begin{eqnarray}
\Theta^{(u)}_\pm (m^2, m_u^2) = 2\,\frac{ p_a(s)\,p_b(s)}{\pi}\,\varrho^{(u)}_\pm (m^2, m_u^2)\, \Bigg|_{s = c^{(u)}_\pm (m^2)}\,,
\label{def-Theta-signature-u}
\end{eqnarray}
are shown. There are integer numbers depending on $m^2$ and $m^2_u$. Like for our t-channel exchange studies the  
plots of the spectral signatures include solid lines that show the full signature with respect 
to  (\ref{def-analytic-continuation-generic-b}) and dotted lines that correspond to the partial expressions (\ref{central-result}). 
Possible contributions at $m^2< 0$ are not shown for the clarity of the presentation. 
In all plots the relevance of a critical point is indicated by its label number as introduced in Tab. \ref{tab:critical-points-u}. 

\def\phx{\phantom{xx}}
\def\phm{\phantom{$-$}}
\begin{table}
\begin{tabular}{l|cccccc}
\hline\hline
&&\phx$K  \,K^*\to K \, K^*$\phx &\phx $\pi \,K\to\pi \,K$\phx &\phx$\pi \,\rho\to \bar{K} \,K^*$ \phx
&\phx$\pi \,K^*\to K\,\rho$\phx &\phx$\pi \,J/\psi\to\pi\,\pi$\\
 \hline\hline
 1 & $u^+_+$ & 8.32172 & 6.71244 & \phm31.9405 & \phm22.2027 & \phm251.181 \\
 2 & $u^+_-$ & 101.333 & 21.0758 & \phm91.8556 & \phm135.925 & $\infty$  \\
 3 & $u^-_+$ & 8.32172 & 6.71244 & $-$2.87974 & \phm5.87230  & $-$21.4357 \\
 4 & $u^-_-$ & 101.333 & 21.0758 & \phm53.8542 & \phm40.9191 & \phm23.4357 \\
 5 & $u_I^+$ & 54.8272 & 13.8941 & \phm44.4880 & \phm70.8985 & $\infty$  \\
 6 & $u_I^-$ & 54.8272 & 13.8941 & \phm42.8973 & \phm31.5609 & \phm1 \\
\hline
 7 & $v^+_+$ & 101.333 & 21.0758 & \phm84.0703 & \phm101.333 & \phm549.234 \\
 8 & $v^+_-$ & 8.32172 & 6.71244 & \phm29.9819 & \phm20.9594 & \phm459.491 \\
 9 & $v^-_+$ & 101.333 & 21.0758 & \phm55.8842 & \phm43.2720  & \phm4 \\
10 & $v^-_-$ & 8.32172 & 6.71244 & \phm3.94940 & \phm8.32172 & \phm0 \\
\hline
11 & $\bar{v}^+_+$ & 101.333 & 21.0758 & \phm145.925 & $-$270.515 & $-$21.4357 \\
12 & $\bar{v}^+_-$ & 8.32172 & 6.71244 & \phm37.4141 & \phm25.9791& \phm23.4357 \\
13 & $\bar{v}^-_+$ & 101.333 & 21.0758 & \phm48.2445 & \phm35.0822& \phm251.181 \\
14 & $\bar{v}^-_-$ & 8.32172 & 6.71244 & $-$46.4972 & $-$5.11443 & $\infty$  \\
\hline
15\; & $u_0$& 109.654 & 27.7882 & \phm87.3853 & \phm102.459 & $\infty$ \\
\hline\hline
\end{tabular}
\caption{Critical points for the u-channel exchange processes shown in Fig.~\ref{fig:diag-list-u}, in units $m_\pi^2$}
\label{tab:critical-points-u}
\end{table}

Consider the first u-channel reaction $ K \, K^* \to  K \, K^*$ of Fig. ~\ref{fig:diag-list-u} and Fig.~\ref{fig:u-channel:A}. 
With $u^-_I > 0$ and 
$v^+_- < v^-_+$ the case  2 in (\ref{central-result}) is selected. This is our first case with $ u^-_+ > 0$. As a consequence both 
contour lines pass through the threshold and pseudo-threshold of this reaction, i.e. $(m_K \pm m_{K^*})^2$. 
This occurs at the critical points $m^2= u^+_+ = u^-_+ > m_\pi^2$ and $m^2=u^+_- = u^-_-> m_\pi^2$.
Since in this reaction a $\pi$-meson is exchanged the relevant parts of the contours do reach both threshold 
points. After all it holds $u^-_+ > m_\pi^2$. Therefore it appears that the extra terms  (\ref{analytic-continuation-result-a}), properly 
transformed from the t-channel kinematics to the u-channel kinematics with $t \leftrightarrow u $, are active in 
this case. The associated closed contour is characterized by $\bar v^-_-< m^2 <v^+_-$ and $v^-_- < m^2 < \bar v^+_-$. However, since 
$\bar v^-_- =v^+_-$ and $ v^-_- = \bar v^+_- $ these are empty conditions for the given example and 
none of the terms in (\ref{analytic-continuation-result-a}) are relevant. On the other hand we may find a contribution of  
(\ref{analytic-continuation-result-b}) since $u^-_- > m_\pi^2$. Here the conditions for the closed contour are 
$\bar v^-_+< m^2 <v^+_+$ and $v^-_+ < m^2 < {\rm Max}( \bar v^+_+, v^+_+) $. Since $\bar v^+_- =v^-_-$ and 
$\bar v^-_- = {\rm Min} (v^+_-, v^-_+) $ again these are empty conditions. Correspondingly, there is no effect of the extra 
terms  (\ref{analytic-continuation-result-b}) also. 

Inspecting the contour paths in Fig.~\ref{fig:u-channel:A} one may be led to the conclusion that the corresponding partial-wave projected 
amplitude has a branch cut going through the two threshold points. However, this is not so.
The effect of the $+$ and $-$ contours in (\ref{disp-general-u-t-channel})
cancel in part, so that the full contribution does not have such a branch cut. Nevertheless, a branch cut emerges  on the real axis, however, 
only at energies where the $+$ and $-$ contours do not overlap. This follows since the $+$ and $-$ spectral signatures have opposite sign. 
As a consequence there is a branch cut connecting the two particular points $c^{(u)}_+(m_\pi^2)> (m_K + m_{K^*})^2$ 
and $c^{(u)}_-(m_\pi^2)> (m_K + m_{K^*})^2$. Here we have an example  where a left-hand branch cut is located right to the largest 
threshold pillar. 

We discuss the $\pi \,K \to \pi\, K$ process of Fig.~\ref{fig:u-channel:A}. This is a further example with $ u^-_+ > 0$ where 
the case 2 in (\ref{central-result}) is scrutinized. Again both 
contour lines pass through the threshold and pseudo-threshold of this reaction, i.e. $(m_\pi \pm m_K)^2$. This occurs at the critical 
points $m^2= u^+_+ = u^-_+ $ and $m^2=u^+_- = u^-_-$. Since in this reaction a $K^*$ meson is exchanged, with a mass distribution 
starting at $ (m_\pi + m_K)^2 > u^-_- > u^-_+ $, the relevant parts of the contours do not reach any of the threshold 
points, however. Like in our first example, even though we have $u^-_- > m_\pi^2$ and $ u^-_+ > 0$, there are no contributions 
from (\ref{analytic-continuation-result-a}) and (\ref{analytic-continuation-result-b}). This is so independent on the value of the 
exchange mass $m_u$. In Fig. \ref{fig:u-channel:A} the contour lines are shown for $m> m_\pi$, in order to illustrate the generic mechanism.

\clearpage

There are the remaining u-channel exchange processes analyzed in Fig. \ref{fig:u-channel:B}. The first two reactions 
$\pi \,\rho \to \bar K\, K^* $ and $\pi \, K^* \to K \,\rho $ probe the case 2 in (\ref{central-result}) with $v^+_- < v^-_+$. The extra 
terms in (\ref{analytic-continuation-result-a}) prove relevant for the second reaction with $u^-_+ > m_\pi^2$ only. The corresponding 
closed contour path is generated by the condition $\bar v^-_-< m^2 < v^+_-$ and $v^-_-< m^2 < \bar v^+_-$ for the minus and plus 
contours respectively (see (\ref{def-closed-contour})). In contrast the additional terms (\ref{analytic-continuation-result-b}) are 
needed in both cases. For the reaction $\pi \,\rho \to \bar K\, K^* $ 
it holds $u^-_- > m_K^2$ and the closed contour is given by $\bar v^-_+ < m^2 < v^+_+$ and $ v^-_+ < m^2 <  \bar v^+_+$. 
A slightly different condition is derived for the $\pi \, K^* \to K \,\rho $ reaction with $u^-_- > m_\pi^2$. 
Here the closed contour follows from $\bar v^-_+ < m^2 < v^+_+$ and $ v^-_+ < m^2 $ instead. 

We discuss  the final u-channel reaction $\pi \,J/\psi\to\pi\,\pi$. It is described by 
the case 4 in (\ref{central-result}) with $v^+_- > v^-_+$. Since it holds $u^-_+ < 0$ here the extra terms 
(\ref{analytic-continuation-result-a}) are not active. On the other hand with $ u^+_+ > m_\rho^2$ the 
terms (\ref{analytic-continuation-result-b}) are needed. The corresponding closed contour follows with
$\bar v^+_- < m^2 < v^+_+$ and $ v^+_- < m^2 $. 

A concluding remark on the numerical implementation of (\ref{disp-general-u-t-channel}) is in order here. 
A partial cancellation of the $+$ and $-$ contour contributions in (\ref{disp-general-u-t-channel}) occurs frequently. 
Whenever the two contours run along identical regions on the real axis this may happen. 
In a numerical implementation of (\ref{disp-general-u-t-channel}) it 
is useful to work out such cancellations explicitly. Based on our general results this is straight forwardly achieved in a 
computer code.  

\newpage 

\section{Summary}

We have analyzed the generic structure of partial-wave projected t- and u-channel exchange diagrams. A general and 
explicit form for a dispersion-integral representation for their contributions to partial-wave reaction amplitudes was established. 
Our results hold for the case of overlapping left- and right-hand cut structures, decaying particles and anomalous 
thresholds or pseudo-thresholds. Various applications to specific examples were worked out and illustrated in detail. 

With our study more realistic treatments of final state interactions in the resonance region of QCD may become feasible. 
The merit of the result lies in its generality. It is a convenient basis for coupled-channel theories 
with a large number of channels involved, where a case-by-case study is prohibitive. 

\vskip1cm
\noindent{\bf Acknowledgments}
\vskip0.5cm
\noindent
M.F.M. Lutz thanks J. Hofmann for collaboration at an early stage of the project. 
C.L. Korpa was partially supported by the Hungarian OTKA fund K109462. 
E.E. Kolomeitsev was supported by the Slovak Grants No. APVV-0050-11 and VEGA-1/0469/15

\clearpage



\bibliography{bibtex}

\appendix

\end{document}